\documentclass[aps,prb,reprint]{revtex4-1}
\usepackage{CJK}
\usepackage{graphicx}
\usepackage{hyperref}
\usepackage{amsmath}

\begin{document}
\begin{CJK*}{UTF8}{gbsn}
\title{Enhanced superconductivity at the interface of W/Sr$_{2}$RuO$_{4}$ point contact}
\author{He Wang (王贺)}
\author{Weijian Lou (娄伟坚)}
\author{Jiawei Luo (骆佳伟)}
\author{ Jian Wei (危健)}\email{weijian6791@pku.edu.cn}
\affiliation{International Center for Quantum Materials, School of Physics, Peking University, Beijing 100871, China}
\affiliation{Collaborative Innovation Center of Quantum Matter, Beijing, China}
\author{Y. Liu}
\affiliation{ Department of Physics and Materials Research Institute, The Pennsylvania State University, University Park, Pennsylvania 16802, USA}
\affiliation{Department of Physics and Astronomy and Key Laboratory of Artificial Structures and Quantum Control (Ministry of Education), Shanghai Jiao Tong University, Shanghai 200240, China}
\author{J.E. Ortmann}
\author{Z.Q. Mao}
\affiliation{Department of Physics, Tulane University, New Orleans, Louisiana 70118, USA}
\date{\today}

\begin{abstract}
Differential resistance measurements  are conducted for point contacts (PCs) between tungsten tip  approaching along the $c$ axis direction and the $ab$ plane of Sr$_{2}$RuO$_{4}$ single crystal. Three key features are found. Firstly, within 0.2 mV there is a dome like conductance enhancement due to Andreev reflection at the normal-superconducting interface.  By pushing the W tip further, the conductance enhancement increases from 3\% to more than 20\%, much larger than that was previously reported, probably due to the pressure exerted by the tip. Secondly, there are also superconducting like features at bias higher than 0.2 mV which persists up to 6.2 K, resembling the enhanced superconductivity under uniaxial pressure for bulk Sr$_{2}$RuO$_{4}$ crystals but more pronounced here. Third, the logarithmic background can be fitted with the Altshuler-Aronov theory of tunneling into quasi two dimensional electron system, consistent with the highly anisotropic electronic system in Sr$_{2}$RuO$_{4}$. 

\end{abstract}
\maketitle
\end{CJK*}

The layered perovskite ruthenate Sr$_{2}$RuO$_{4}$ (SRO) has shown evidence for spin-triplet, odd-parity superconductivity (SC) which may be useful for topological quantum computation.~\cite{Maeno1994nature,Mackenzie2003rmp,Maeno2012jpsj} The possible chiral orbital order parameter for the two-dimensional SC is $p_{x}\pm ip_{y}$ as suggested by the time-reversal symmetry breaking experiments.~\cite{Luke1998nature,Xia2006prl} Such chiral order is expected to generate edge currents, but the expected magnetic field due to edge currents has not been directly observed with local field imaging,~\cite{Kirtley2007prb,Hicks2010prb,Curran2014prb} though there is indirect evidence of edge currents revealed by in-plane tunneling spectroscopy~\cite{Kashiwaya2011prl,Kashiwaya2014pe} and point contact spectroscopy (PCS),~\cite{Laube2000prl}both with assumptions to fit the conductance spectra. 

The surface properties of  SRO is very critical for field imaging with scanning quantum interference devices, as well as for the tunneling and point contact spectroscopy. It is known that the SRO surface can undergo reconstruction and the intrinsic SC may not be probed,~\cite{Upward2002prb,Firmo2013prb} and it may even show ferromagnetism (FM) due to lattice distortion.~\cite{Matzdorf2000science} Very careful \textit{in situ} preparation of devices is required for making good  tunnel junctions using microfabrication techniques.~\cite{Kashiwaya2011prl} Recently there is also theory proposal that surface disorder indeed can destroy the spontaneous currents.~\cite{Lederer2014prb} 

One way to overcome the surface problem is to use a hard tip for the point contact (PC) measurement. If the tip is hard enough, it may pierce through the surface dead layer and probe the SC underneath.~\cite{Gonnelli2002jpcs} In fact, for this reason tungsten tip has been used for PCS of heavy fermion superconductors.~\cite{Gloos1996jltp_scaling} A consequence of using a hard tip is that the tip will exert some pressure on the surface which may affect the SC,~\cite{Daghero2010sst} possibly due to local distortion of lattice.~\cite{Gloos1995pb,Miyoshi2005prb} It is known that for SRO a very low uniaxial pressure of 0.2 GPa along the $c$ axis can enhance the superconducting transition temperature ($T_c$) of pure SRO from 1.5 K up to 3.2 K,~\cite{Kittaka2010prb,Kittaka2009jpsj_b} and recently in-plane strain (0.23\%) along $\langle 100 \rangle$ direction is also shown to enhance $T_c$ from 1.3 K up to 1.9 K.~\cite{Hicks2014science} The pressure in abovementioned measurements were applied to bulk samples, while for PCS the pressure is exerted locally. In the latter case it may be less affected by the inhomogenity of the applied pressure and the sample is less tend to developing cracks, thus locally higher pressure may be reached though absolute pressure is not known. Here we report greatly enhanced SC observed at the interface of the point contact junction between a tungsten tip approaching along the $c$ axis direction and the $ab$ plane surface of a SRO single crystal.

SRO single crystals are grown by floating zone methods and are from two different batches, details of sample preparation can be found in previous reports.~\cite{Mao2000mrb} Sample S1 is from the first batch and is easier to cleave and shows no Ru inclusions. Sample S2 is from the second batch, too hard to cleave, and contains a lot of Ru inclusions (for optical images see  Appendix~\ref{appendix_Ru_inclusions}). Only on the   cleaved surface of S1 do we observe SC feature. Tungsten wire of 0.25 mm diameter is etched to form the tip, and then fixed pointing to the $ab$ plane of the SRO sample. A Si chip with the sample and thermometer glued on top is mounted on an attoCube nanopositioner stack. Since the tip and sample are both fixed to the copper housing, relative displacement between the tip and sample is suppressed, which ensures a stable contact and reproducible PCS. The housing is suspended with springs at the bottom of a insertable probe for a Leiden dilution fridge. With such customization  the sample position is not at the field center of the magnet, and the field value is estimated with the tabled values from the magnet manufacturer. Differential resistance ($dV/dI$) is measured with standard lock-in technique.

\begin{figure}
\includegraphics[width=9cm]{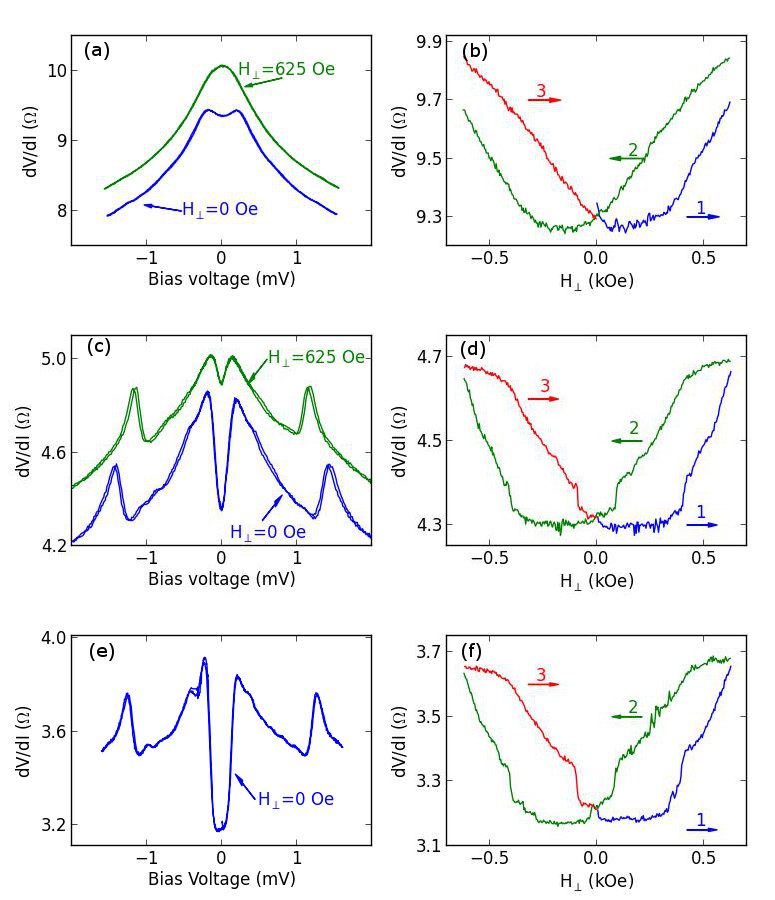}
\caption  {\small (Color online) Bias dependence of $dV/dI$ (a, c, e) and magnetoresistance (b, d, f) of three different point contact (PC) resistance at the same location between the W tip and SRO single crystal S1 at 0.35 K. The resistance at zero bias and zero field is 9.3, 4.3, 3.2  $\Omega$ respectively. For clarity,  in (a) and (c) the $dV/dI$ curves at 625 Oe (Green) are shifted up by 0.2 $\Omega$. Arrows in (b), (d), (f) show the sweeping direction of the magnetic field. The reproducibility of the measurements is demonstrated by the overlapping of $dV/dI$ curves in (a), (c), (e) with bias ramping in both directions. The discontinuity around $\pm$625 Oe is related to the ramping speed of the field, and can be smaller when the field ramping speed is reduced, while the hysteresis is almost the same.} 
\label{fig_dVdI_pressure}
\end{figure}

At the same location, by pushing the tungsten tip towards the SRO surface (more precisely it is the SRO moving towards the tip), the PC resistance is reduced and the pressure is increased. The zero bias and zero field resistance ($R_0$) is: 9.3, 4.3, 3.2  $\Omega$ respectively (see Appendix~\ref{appendix_R_pc} for a discussion of PC resistance). The bias dependence of $dV/dI$ is shown in  Figs.~\ref{fig_dVdI_pressure}a, \ref{fig_dVdI_pressure}c, and \ref{fig_dVdI_pressure}e, at nominal temperature 0.35 K. SC is clearly shown by the resistance dip within $\pm$0.2 mV  without any applied field. With a 625 Oe magnetic field applied along the $c$ axis (H$_{\perp}$), SC is almost fully suppressed for the  9.3 $\Omega$ PC as shown by the recover of the resistance peak at zero bias. However, for the 4.3  $\Omega$ PC there is still a small dip, suggesting that SC is not fully suppressed, \textit{i.e.}, SC is enhanced with increased pressure. 

\begin{figure}
\includegraphics[width=9cm]{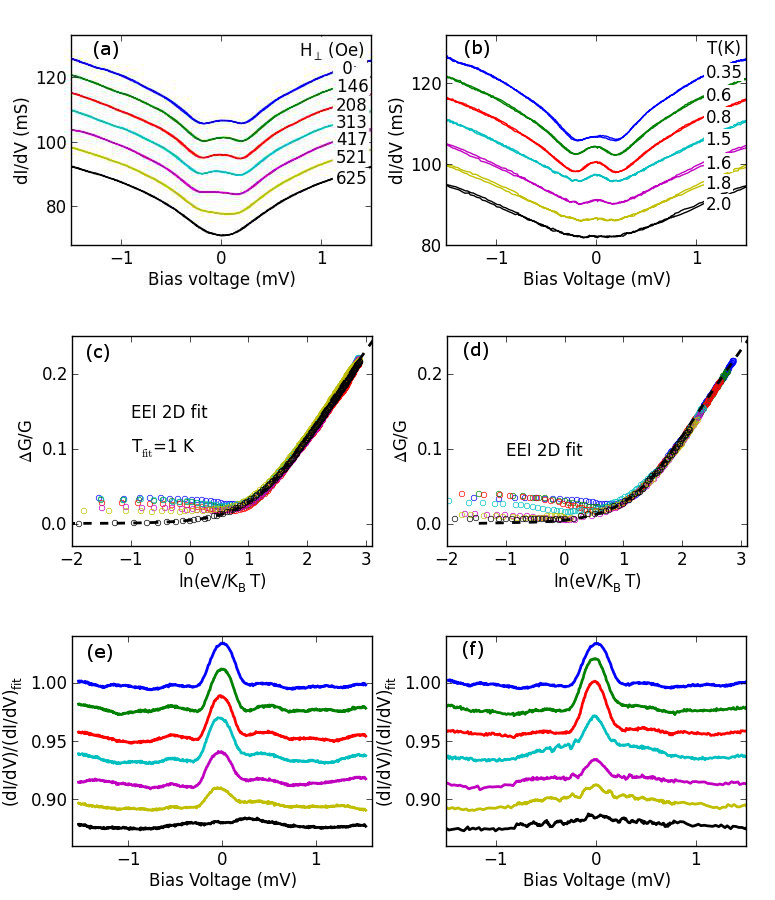}
\caption  {\small (Color online) (a) Bias dependence of $dI/dV$ for the 9.3 $\Omega$ point contact at 0.35 K and with increasing $H_{\bot}$ and (b) zero field $dI/dV$ with increasing T. Curves are shifted for clarity except for the  zero field 0.35 K curve. (c) Fitting with EEI theory in the 2D limit for curves in (a) with fitting temperature $T_{fit}$=1.0 K, and  (d) fitting for curves in (b) with $T=$ 0.35 (1.0), 0.6 (1.1), 0.8 (1.24), 1.5 (1.65), 1.6 (1.85), 1.8 (1.95), and 2.0 (2.2) K from top to bottom ($T_{fit}$ is indicated in the parentheses).  After normalized by  the EEI fits with corresponding $T_{fit}$, the data curves are shown in (e) for different $H_{\bot}$ and (f) for different T. Curves are shifted for clarity.  } 
\label{fig_dVdI_9ohm}
\end{figure}

\begin{figure}
\includegraphics[width=9cm]{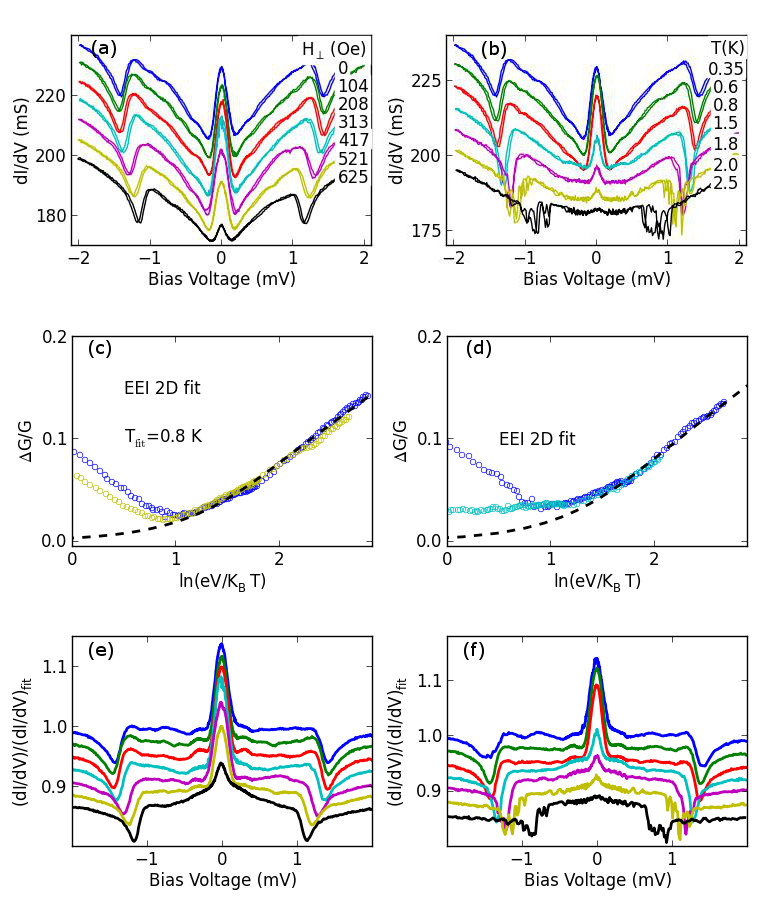}
\caption  {\small (Color online) (a) Bias dependence of $dI/dV$ for the 4.3 $\Omega$ point contact at 0.35 K and with increasing $H_{\bot}$ and (b) zero field $dI/dV$ with increasing T. Curves are shifted for clarity except for the zero field 0.35 K curve. (c) Fitting with EEI theory in the 2D limit for $H_{\bot}$ = 0 (blue) and 625 Oe (yellow) curves in (a) with $T_{fit}$=0.8 K, and  (d) fitting for different T curves in (b) with $T$ = 0.35 (0.8), 1.5 (1.5) K ($T_{fit}$ is indicated in the parentheses).  After normalized by  the EEI fits with corresponding $T_{fit}$, the data curves are shown in (e) for different $H_{\bot}$ and (f) for different T. Curves are shifted for clarity.} 
\label{fig_dVdI_4ohm}
\end{figure}

Enhancement of SC is further confirmed by the temperature dependence of $dI/dV$ at zero field as shown in Fig.~\ref{fig_dVdI_9ohm}b and Fig.~\ref{fig_dVdI_4ohm}b, where $T_c$ is increased from the bulk value of 1.5 k to about 2 K and 2.5 K for the 9.3 $\Omega$ and 4.3  $\Omega$ PC  respectively.  This  enhanced T$_{c}$ is consistent with previous susceptibility measurements on bulk SRO sample under uniaxial pressure, where  the mechanism of $T_c$ enhancement was ascribed to anisotropic lattice distortion,~\cite{Kittaka2009jpsj_b,Kittaka2010prb,Taniguchi2012jpcs} similar to that found in the eutectic 3K phase.~\cite{Ying2009prl,Ying2013ncomms} In Figs.~\ref{fig_dVdI_9ohm} and ~\ref{fig_dVdI_4ohm}, for easy comparison with theoretical description, $dV/dI$ is converted to $dI/dV$.

The magnetoresistance (MR) is shown in Figs.~\ref{fig_dVdI_pressure}b, \ref{fig_dVdI_pressure}d, and \ref{fig_dVdI_pressure}f for the three PCs. The resistance starts to increase quickly at around 400 Oe, and there is clearly a hysteresis with steps which gets sharper and more pronounced for higher PC pressure. MR hysteresis is usually observed for ferromagnetic samples, and the observation of both SC and MR hysteresis was linked to the coexistence of SC and ferromagnetism (FM) for SC at the oxides interface.~\cite{Dikin2011prl} If indeed a FM-like internal field exists, could it be related to the long sought-after time-reversal symmetry-breaking fields?~\cite{Kirtley2007prb,Hicks2010prb,Curran2014prb} 

First the possibility of conventional vortex pinning needs be considered.  The field value above which $dV/dI$ starts to increase quickly is around 400 Oe,  in the same order of magnitude with the upper critical field $H_{c2}||c$ about 710 Oe for pure SRO crystal, but much larger than the critical field $H_{c1}||c$ about 70 Oe (by specific heat measurements).~\cite{Deguchi2004jpsj} Sharp increase of resistance may indicate that vortices enter the PC interface and SC is suppressed. However, it is not clear whether such strong pinning could be reduced by PC. The average distance between vortices is  $\sim\sqrt{\Phi_{0}/H}$, about 0.3 $\mu$m for 400 Oe, so the diameter of the PC should be much larger to include multiple vortices, which is inconsistent with conventional understanding of the PC. Moreover, it is difficult to explain why the step-like features become sharper with higher pressure. 
Besides the external pinning due to defects, the intrinsic pinning due to chiral domain wall~\cite{Dumont2002prb} seems also unlikely to reach 400 Oe.
One variation of vortex pinning is chiral domain wall motion, where with ramping field the DW wall moves and the edge current can affect the transport of the PC,~\cite{Kambara2008prl} which seems reasonable.

Surface FM also needs to be considered since among other layered perovskite ruthenates in the series $A_{n+1}Ru_{n}O_{3n+1}$, SrRuO$_{3}$ is a ferromagnetic metal with T$_{c}$=160 K, and Sr$_{3}$Ru$_{2}$O$_{7}$ is at the boarder of FM and shows pressure-induced FM.~\cite{Ikeda2000prb} Thus it is natural to expect that FM could be induced for SRO, or there might be some eutectic phase~\cite{Mao2000mrb} on the surface which leads to FM. Previously, experimental attempts to measure the bulk magnetic susceptibility of SRO with uniaxial pressure were not successful, since above 0.4 Gpa SRO sample tends to crush,~\cite{Ikeda2004jmmm} while no drastic change of the temperature dependence of susceptibility was observed. On the other hand, doping the Sr with Ca does show a ground state of \textit{static} magnetic order due to rotation of $RuO_{6}$ octahedra.~\cite{Carlo2012nmat,Ortmann2013sr} Thus it is possible that the pressure under the tip may be higher than 0.4 GPa~\cite{Gonnelli2002jpcs} and its influence is comparable with that by doping. However, this is inconsistent with the fact that the hysteresis  diminishes together with SC at higher temperatures, which also indicates that the hysteresis is not due to eutectic phase impurities. 

Both field and temperature dependences of $dI/dV$ resemble those found for in-plane Au/SRO tunneling junctions in Ref.~[\onlinecite{Kashiwaya2011prl}], as shown by detailed field and temperature dependences in Fig.~\ref{fig_dVdI_9ohm} and Fig.~\ref{fig_dVdI_4ohm}, for the 9.3 and 4.3 $\Omega$ PC  respectively. 
However, in Ref.~[\onlinecite{Kashiwaya2011prl}] the gap is about 0.7 mV instead of 0.2 mV, 
and the conductance enhancement of the dome like feature is less than 1\% (see see Appendix~\ref{appendix_reproducibility} for similar PC spectra with a Au tip). The dome like feature may be fitted considering chiral \textit{p}-wave symmetry~\cite{Kashiwaya2011prl}, but here we focus on experimental findings and methodology while leave the fittings in the future.

\begin{figure}
\includegraphics[width=9cm]{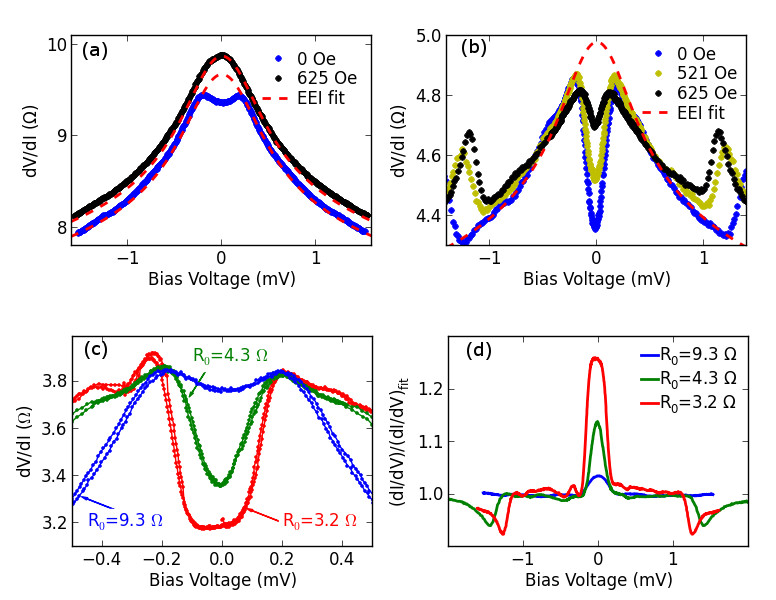}
\caption  {\small (Color online) (a) $dV/dI$ curves  for the 9.3 $\Omega$ PC at 0 (blue), 625 Oe (black), and EEI fits (red dashed lines). (b) $dV/dI$ curves  for the 4.3 $\Omega$ PC at 0 (blue), 521 (yellow), 625 Oe (black), and the EEI fit. (c) Zoom-in of the zero bias resistance dip regime with curves shifted for clarity except for the 3.2  $\Omega$ curve. The curves are reproducible for both ramping directions of the bias. (d) Zero field conductance enhancement after normalized with the fitted EEI  background. All at 0.35 K.} 
\label{fig_dVdI_difference}
\end{figure}

The broad  background resistance hump as shown by $dV/dI$ at 625 Oe in Fig.~\ref{fig_dVdI_difference}(a) (same as in Fig.~\ref{fig_dVdI_pressure}a) is generally called zero-bias anomaly (ZBA), which is frequently observed in tunnel junctions~\cite{Kashiwaya2011prl} as well as PCs.~\cite{Gloos2009ltp,Gloos2012jpcs} The possible origins for ZBA in PCs include ``extrinsic'' magnetic impurities, two-level systems, Kondo scattering due to spontaneous electron spin polarization etc, as well as ``intrinsic''  density of states (DOS) effect, as shown for chromium where DOS is reduced due to the spin density wave gap,~\cite{Meekes1988prb} and more recently for iron pnictides where DOS is enhanced due to strong electron correlations.~\cite{Arham2012prb} Here ZBA apparently coexists with SC in SRO, which is very sensitive to impurities, thus the origin of ZBA is more likely due to some ``intrinsic'' origin.

The background ZBA can be normalized when the bias dependence is replotted using $\ln{(eV/k_{B}T)}$. In Figs.~\ref{fig_dVdI_9ohm}(c) and ~\ref{fig_dVdI_9ohm}(d), the normalized change of conductance shows a linear dependence for $eV \gg k_{B}T$, similar to what was observed in tunneling measurements for disordered metal films,~\cite{Gershenzon1986jetp} and also for layered cuprates and manganites.~\cite{Abrikosov2000prb,Mazur2007prb} In the tunneling case, the reduction of DOS is due to electron-electron interaction (EEI). As proposed by Altshuler and Aronov,~\cite{Altshuler1979,*Altshuler1980,*Altshuler1985} for low dimensional systems the exchange interaction between electrons can cause quantum corrections to the conductivity as well as DOS, which depends on the dimensionality of the systems. For $eV \gg k_{B}T$, the DOS correction $\sim\ln{(eV/k_{B}T)}$ in 2D (see Appendix~\ref{appendix_fit_eei} for details). When the full formula is used, we get good fits in the full bias range as shown by the dashed lines in Figs.~\ref{fig_dVdI_9ohm}(c) and ~\ref{fig_dVdI_9ohm}(d) (also in Figs.~\ref{fig_dVdI_pressure}a and b).
We note that in order for all normalized $dI/dV$ curves to collapse onto a single curve,  enhanced temperature ($T_{fit}$) needs to be assumed for $dI/dV$ measured at lower temperatures. This may indicate there is local heating in the small PC region, possibly due to inadequate filtering of the external microwave noise.~\cite{Liu2014prb}

The fitted ZBA can be considered as the normal state background and divided from the normalized conductance,~\cite{Kashiwaya2011prl} the resulted curves are shown in Figs.~\ref{fig_dVdI_9ohm}(e) and ~\ref{fig_dVdI_9ohm}(f), with the dome like conductance enhancement well demonstrated. Another feature of the PC spectra is a small periodic ``wiggling'' outside $\pm$0.2 mV, which also diminishes with increasing field and temperature,  suggesting that it is probably due to interference of quasiparticles at the NS interface. Similar feature was also observed for multiple band superconductor MgB$_2$~\cite{Gonnelli2002jpcs,Daghero2010sst} but detailed analyses are lacking.
The ZBA background becomes less pronounced when the PC resistance is reduced from 9.3 $\Omega$ to 4.3  $\Omega$, as shown in Figs.~\ref{fig_dVdI_4ohm}c and d, while the conductance enhancement gets larger. This is better illustrated by the normalized enhancement (Fig.~\ref{fig_dVdI_difference}d), and by direct comparison of the zero field $dV/dI$ (Fig.~\ref{fig_dVdI_difference}c). 

What parameters may change when the PC resistance is reduced from 9.3 $\Omega$ to 4.3  $\Omega$? In the standard theory for PCS (see Appendix~\ref{appendix_R_pc} for details), the PC  resistance 
 \begin{equation}
R_{PC}=R_{Sh}+R_{Max},
\label{Eq_R_pc}
\end{equation}
where $R_{Sh}$ is the Sharvin resistance corresponding to the ballistic limit, and $R_{Max}$ is the Maxwell resistance corresponding to the diffusive limit and related to the resistivity. For the simplest metallic PC, $R_{Sh}$ is considered to be energy independent as the energy dependence of velocity cancels that of the DOS. This can be changed when complicated Fremi surface is involved and the effective DOS may be probed by $R_{Sh}$.~\cite{Meekes1988prb,Arham2012prb} Here for single crystal SRO the mean free path is large, and if the interface is clean and barrier-free, the PC should be close to the ballistic limit. As $R_{Sh} \propto (1/d)^2$ and $R_{Max} \propto (1/d)$, where $d$ is the diameter of PC, and if the anisotropic electronic state in SRO is not considered,  the reduction of resistance from 9.3 $\Omega$ to 4.3  $\Omega$ would lead to an increase of $d$ by roughly $\sqrt{9.3/4.3}$=1.47 times in the ballistic limit (twice increase of the area); or  by 2 times in the thermal limit(quadruple increase of the area).

\begin{figure}
\includegraphics[width=9cm]{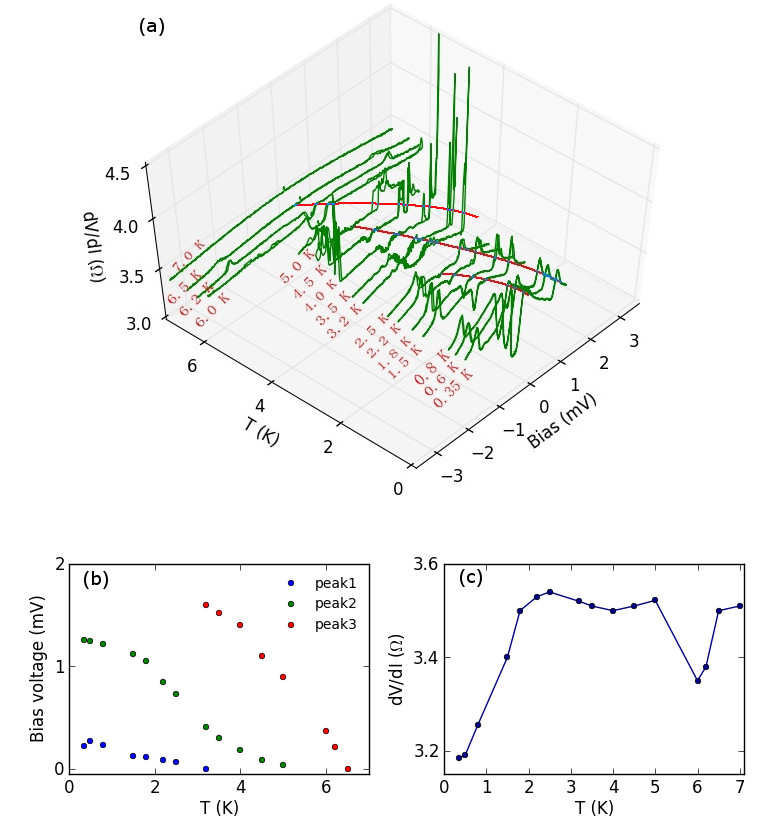}
\caption  {\small (Color online) (a) For $R_0$=3.2 $\Omega$ three SC transitions are shown by $dV/dI$ curves of  at different temperatures. Red curves are guide to the eye. (b) The position of the $dV/dI$ peaks vs temperature. (c) Zero bias $dV/dI$ vs temperature. Both (b) and (c) are derived from (a).} 
\label{fig_dVdI_3ohm}
\end{figure}

With the increase of contact area, the PC may show a larger critical current ($I_C$) if the critical current density is constant and $I_C$ is \textit{only} determined by the PC itself. As the additional $dI/dV$ dips shown in Fig.~\ref{fig_dVdI_4ohm} is ascribed to the critical current effect~\cite{Sheet2004prb}, $I_C$ can be estimate from the dip position. At 1.6 K the dip position is about 1.2 and 2.3 mV for the 4.3  $\Omega$ and 9.3  $\Omega$ PCs,  so the calculated $I_C$ is around 0.28 and 0.25 mA respectively, inconsistent with the expected 2-4 times increase of $I_C$ if $I_C$ is proportional to the contact area. This may suggest that $I_C$  is determined by a fixed region, e.g., chiral domains under the PC, instead of by the area of the PC itself. Thus, with increasing bias the region of SRO under the PC reaches its $I_C$, and $R_{PC}$ shows a finite increase due to $R_{Max}$, as described in Eq.(\ref{Eq_R_pc}). 

When the PC resistance is reduced further to 3.2 $\Omega$, even larger conductance enhancement is observed as shown in Fig.~\ref{fig_dVdI_pressure}e and Fig.~\ref{fig_dVdI_difference}.  
After normalization by the background, the conductance enhancement at zero bias is about 3\% for the 9.3 $\Omega$ PC, 14\% for the  4.3 $\Omega$ PC, and 22\% for the 3.2 $\Omega$ PC (Fig.~\ref{fig_dVdI_difference}d). 
The original $dV/dI$ curves without normalization and the EEI fits are also shown in Figs.~\ref{fig_dVdI_difference}a and b,  and the fits can very well reproduce the $dV/dI$ curves when an effective temperature $T_{fit}$ is taken into account.
Zoom-in the zero bias regime, the absolute amplitude of the $dV/dI$ dip and of the ``wiggling'' part outside of the dip are clearly shown in Fig.~\ref{fig_dVdI_difference}c.
For all three PCs, the $dV/dI$ dip evolves to ZBA at around 0.2 mV (Fig.~\ref{fig_dVdI_difference}c), which is consistent with the gap value of SRO from the weak-coupling theory ($2\Delta/e=3.5k_{B}T_{c}$) with $T_{c}\sim 1.5$ K. This value is much smaller compared with previous PCS and $ab$ plane tunneling results where 0.7-0.9 mV were obtained,~\cite{Laube2000prl,Kashiwaya2011prl} and also smaller compared with scanning tunnel spectroscopy (STS) measurements where 0.3-0.5 mV were reported.~\cite{Upward2002prb,Firmo2013prb} 

For the 3.2 $\Omega$ PC, the surprising feature that the critical current effect  persists to much higher temperatures is better illustrated in Fig.~\ref{fig_dVdI_3ohm}.  In Fig.~\ref{fig_dVdI_3ohm}a, besides the first $dV/dI$ peak at around 0.2 mV, there are two additional $dV/dI$ peaks, one persists up to about 5 K, while the other persists up to about 6.2 K. These two $dV/dI$ peaks are likely to be SC features as the measured MR up to around 625 Oe also shows hysteresis, which decreases with increasing temperature and diminishes along with the resistance dip near zero bias.  At 8 K, the $dV/dI$ within $\pm$1.5 mV, and the MR of the zero bias $dV/dI$  within $\pm$625 Oe, becomes practically flat and changes less  than 0.03 $\Omega$. The temperature dependence of the position of $dV/dI$ peaks is plotted in Fig.~\ref{fig_dVdI_3ohm}b, and the zero bias $dV/dI$ from the spectra in Fig.~\ref{fig_dVdI_3ohm}a is plotted in Fig.~\ref{fig_dVdI_3ohm}c. There are clearly two resistance drops at around 4 and 6 K, and we note a similar but smaller drop around 4 k was also observed in Ref.~[\onlinecite{Kashiwaya2011prl}]. Since the bulk $T_{c}\sim$ 1.5 K for S1, and even for the 3-K phase $T_{c}\sim$ 3 K, thus the greatly enhanced $T_{c}$ could be only due to the W/SRO PC.

In summary, an ultralow temperature point contact setup using nanopositioners was used to measure differential resistance of W/SRO point contact junctions. We find: 1) a superconducting gap around 0.2 mV and a dome like shape of conductance enhancement, consistent with chiral \textit{p}-wave symmetry; 2) SC-like features persisting up to 6.2 K, much higher than the bulk T$_c$ of SRO, presumably due to the pressure exerted by the W tip and a mechanism similar to that of the 3K-phase; 3) a broad   resistance hump coexisting with superconductivity, which is ascribed to density of states effect due to 2D electron-electron interaction, consistent with the highly anisotropic electronic system of SRO. We believe PCS may provide useful information beyond the surface problem for SRO.

We thank Hu JIN for contribution in the earlier stage of this project, Liang LIU for various help with experiments and data analysis, Xin LU for helpful discussions on point contact measurements, and Fa WANG for discussions on correlated systems respectively. Work at Peking University is supported by National Basic Research Program of China (973 Program) through Grant No. 2011CBA00106 and No. 2012CB927400. The work at Tulane is supported by the DOE under grant DE-SC0012432. 


\appendix

\section{Basics of point contact resistance}
\label{appendix_R_pc}

There are many reviews on point contact spectroscopy~\cite{Duif1989jpcm} and particular on unconventional heavy fermion systems.~\cite{Naidyuk1998jpcm,Park2009jpcm} Here we introduce the basics of the PC resistance following Ref.~\onlinecite{Naidyuk1998jpcm}.

In the simple theoretical model, PC is formed with an orifice with diameter $d$ between two bulk metallic electrodes. Depending on the relative ratio between $d$ and different mean free path $l$, PC can be categorized into three regimes: ballistic ($d<l_{elastic}$), diffusive ($l_{elastic}<d<l_{inelastic}$), and thermal ($d>l_{inelastic}$). In the ballistic regime, the Fermi surface in the two electrodes has a difference of $eV$, similar to the tunneling junction case; while in the thermal regime, the Fermi surface evolves smoothly within the PC and there is a well defined equilibrium temperature profile.~\cite{Pothier1997} 

The current density in the orifice along its normal direction ($z$-axis) is 
\begin{equation}
j_{z}=2e \sum_{k}(v_{k})_{z}f_{k}(E),
 \end{equation}
where $v_{k}$ is the electron velocity, and $f_{k}(E)$ is the Fermi-Dirac distribution function.
For a voltage biased ballistic PC, considering the energy difference $eV$,
\begin{equation}
j_{z}=e \int_{E_{F}-eV/2}^{E_{F}+eV/2}dE \int \frac{d\Omega}{4\pi} v_{z}(E)f(E)N(E),
 \end{equation}
where $N(E)$ is the electronic DOS. In the simplified case, $v_{z}(E)$ is inversely proportional to $N(E)$, thus there is no non-linearity caused by energy dependence of DOS. The resulted Ohmic resistance is 
\begin{equation}
R_{Sh}=\frac{16R_{q}}{(k_{F}d)^{2}}=\frac{16\rho l}{3\pi d^{2}},
\end{equation}
where $\rho$ is the bulk resistivity, $l$ the elastic mean free path, $R_{q}=h/2e^{2}=12.9$ k$\Omega$ the quantum resistance. 
With the assumption that the Drude picture holds, $\rho l = p_{F}/ne^{2}$  is a constant for a particular metal (Note that the quantities $p_{F}$ and $n$ were used in the original derivation). Thus, in the ballistic regime the diameter of the orifice $d$ can be estimated using the zero bias resistance $R_{0}$. To get a rough number, in the case of copper and other simple metals, $d\sim 30/\sqrt{R_{0}(\Omega)}$ nm. 

At finite bias, the electron can also be backscattered by phonons, magnons etc, at characteristic bias energy. So $I$-$V$ curve of the ballistic PC can be nonlinear and second derivative is often used to identify phonon and magnon spectra.   More generally, for correlated materials with complex Fermi surface, $v_{z}(E)$ is no longer inversely proportional to $N(E)$, $I$-$V$ curve is nonlinear and $R_{Sh}(E)$ may reflect the change of DOS.~\cite{Meekes1988prb,Arham2012prb}

For PC in the diffusive or thermal regime, electrons in the PC are scattered by impurities or defects, whose contribution to $R_{PC}$ can be estimated from the bulk resistivity, and the orifice just provides a geometric limitation. In the limit $d\gg l_{inelastic}$, the Maxwell resistance is
\begin{equation}
R_{Max}=\frac{\rho}{d}.
\end{equation}
As it depends on $d^{-1}$ instead of $d^{-2}$, it dominates over $R_{Sh}$ when $d$ is large. And when inelastic scattering happens inside the PC, the equilibrium temperature in the PC can be elevated following
\begin{equation}
T_{PC}^{2}=T_{bath}^{2}+\frac{V^{2}}{4L}
\end{equation}
where $L$ is the Lorentz number. For a rough estimation, when $T_{bath}\ll T_{PC}$, assume a standard $L=2.45\times 10^{-8}$ V$^{2}$K$^{-2}$, then $eV\sim 3.63k_{B}T_{PC}$, or $T_{PC}$ (K)$\simeq$ 3.2$V$ (mV). That explains for a thermal PC similar feature can be found in $dV/dI(V)$ and in $dV/dI(T)$. For the gap energy around 0.2 mV in this work, in the thermal limit a rough estimation of $T_{PC}$ at 0.2 mV is 0.64 K, which is below the $T_C$ of SRO, so the bias will not drive the PC out of the SC state even in the thermal limit.

In the intermediate regime, Wexler derived an interpolation formula
\begin{equation}
R_{PC}(T)\simeq\frac{16\rho l}{3\pi d^{2}}+\frac{\rho(T)}{d}.
\end{equation}
For a heterocontact between two different electrodes (1 and 2), the resistance has contribution from both sides. For geometrically symmetric PC with almost equal $p_{F}$, 
\begin{equation}
R_{PC}(T)\simeq\frac{16\rho l}{3\pi d^{2}}+\frac{\rho_{1}(T)+\rho_{2}(T)}{2d}.
\end{equation}

Since the resistivity of simple metal tip like tungsten is usually much smaller than that of the correlated electron systems (in normal state), we may just keep the resistivity term of the correlated systems being probed. The assumption of equal $p_{F}$ is very rough, the difference between $k_F$ of tungsten and SRO is shown in Table~\ref{table_k_F}. Here $k_F$ of tungsten is roughly estimated by assuming two valence electrons and simple spherical Fermi surface. 

\begin{table}
\caption{Summary of quasiparticle parameters of Sr$_2$RuO$_4$ ($\alpha$,$\beta$,$\gamma$)\cite{Mackenzie2003rmp} and Tungsten.}
\begin{tabular}{ccccc}
  \hline
  Fermi sheet & $\alpha$ & $\beta$ & $\gamma$ & Tungsten \\
  \hline\hline
	$k_{F}$ (${\AA}^{-1}$) & 0.304 & 0.622 & 0.753 & 1.55 \\
	$v_{F}$ ($ms^{-1}$) &$1.0\times 10^{5}$ & $1.0\times 10^{5}$ & $5.5\times 10^{4} $& $1.8\times 10^{6} $\\
	$m^{*}$ ($m_{e}$) &3.3 & 7.0 & 16 &1\\
  \hline
\end{tabular}
\label{table_k_F}
\end{table}

For a heterocontact between a normal metal and a superconductor,  Blonder-Tinkham-Klapwijk (BTK) model~\cite{Blonder1982prb} is widely used to explain the conductance enhancement within the gap energy and a tunnel barrier $Z$ parameter is used to characterize the interface. Whether the Fermi velocity mismatch can be represented with an effective $Z$ parameter is not yet clear.~\cite{Park2009jpcm} Note that in BTK model the scattering in the metals and the interface is not considered, even for finite $Z$. So its transparent interface limit ($Z=0$) corresponds to the ballistic limit of the PC model, \textit{i.e.},  the point contact Andreev reflection spectroscopy can only be applied to \textit{ballistic} contacts. Since the BTK model can be used for various interface transparencies, it has wider application than the simple $Z=0$ point contact model. To take into account additional scattering at or near the interface, \textit{i.e.}, $R_{Max}$, a normal resistor in series~\cite{Sheet2004prb} or a normal current in parallel~\cite{Miyoshi2005prb,Peng2013prb} can be added. Thus even in the so-called thermal regime, the gap value can be roughly estimated with consideration of a combination of the BTK model and PC model.~\cite{Sheet2004prb}

In some cases it is believed that although the footprint of the PC can be tens of microns, much larger than $l$, but still ballistic limit can be applied because there are multiple smaller PC junctions randomly distributed across the contact area,~\cite{Bugoslavsky2005prb,Peng2013prb} and the BTK model can be used directly. Although conceptually this is different from the picture that there is an interface barrier which contributes to the PC resistance like a real tunneling junction, but in both cases ballistic limit can be applied as $R_{Max}$ is smaller than $R_{Sh}$.

When the SC has unconventional pairing symmetries, generalized BTK model is developed to fit the data by taking into account various parameters including order parameter symmetry, incidence angle, Fermi surface mismatch, life time broadening due to inter or intra band scattering etc. PCS for unconventional SC has been reviewed in Ref.~\onlinecite{Gloos1996jltp_scaling,Park2009jpcm,Daghero2010sst,Daghero2013ltp}. It is still not clear whether  the order parameter symmetry can be verified strictly from the shape of the point contact Andreev reflection spectra.~\cite{Gloos1996jltp_scaling,Park2009jpcm} In this work we mainly report the temperature and field dependence of the PC spectra rather than quantitatively fit the data with the generalized BTK model.~\cite{Kashiwaya2011prl,Kashiwaya2014pe} 


\section{Fitting with electron-electron interaction}
\label{appendix_fit_eei}

The difference between PCS and planar tunneling is whether  the in-plane momentum is conserved. Since there is no well-known theory for incorporation of quantum correction of DOS into PCS, here we use the theory for the  planar tunneling junctions.

Correction to tunneling conductance by electron-electron interaction (EEI) is quantitatively described by the Altshuler-Aronov (AA) theory,~\cite{Altshuler1985,Gershenzon1986jetp} in the 2D limit, 
\begin{equation}
\frac{G(V,T)-G(0,T)}{G(0,T)}=\frac{e^{2}R_{sq}}{8\pi^{2}\hbar}\ln{\frac{4\pi\delta}{\mathcal{D}R_{sq}}}[\Phi_{2}(\frac{eV}{k_{B}T})-\Phi_{2}(0)],
\label{Eq_AA_2d}
\end{equation}
where  $R_{sq}$ is the resistance per square of the metal film, $\delta$ the thickness of the insulating barrier, $\mathcal{D}$ the diffusion constant, and $\Phi_{2}$ a integral for 2D as defined in Ref.~[\onlinecite{Gershenzon1986jetp}]. The integral is 
\begin{equation}
\begin{aligned}
& \Phi_{d}(A)= \int_{-\infty}^{\infty}dx{\frac{\cosh(x+A)-1}{{\cosh(x/2)}^{2}}} \\
& \times\int_{0}^{\infty}dx\frac{\sinh{y}dy}{[\cosh{y}+\cosh(x+A)](1+\cosh{y}){y^{2-d/2}}},
\end{aligned}
\label{Eq_AA_Phi}
\end{equation}
where $x=\epsilon/kT$ and $A=eV/kT$.

The prefactor before the bracket in Eq.~(\ref{Eq_AA_2d}) can be lumped into one parameter $S$ and it is the only fitting parameter. When $eV\gg k_{B}T$ but still within the 2D limit, Eq.~(\ref{Eq_AA_2d}) approaches $S\ln{\frac{eV}{k_{B}T}}$ and $S$ is just the slope shown in Fig.~2. Since $R_{sq}=\rho/a$, $a$ the thickness of the metal film, the resistivity $\rho=(e^{2}\nu \mathcal{D})^{-1}$, the slope $S \propto R_{sq}\ln(c\nu)$, where $c$ is a constant.

For the 3D limit,
\begin{equation}
\frac{G(V,T)-G(0,T)}{G(0,T)}=\frac{e^{2}\rho}{8\sqrt{2}\pi^{2}\hbar}(\frac{k_{B}T}{\hbar\mathcal{D}})^{1/2}[\Phi_{3}(\frac{eV}{k_{B}T})-\Phi_{3}(0)],
\label{Eq_AA_3d}
\end{equation}
which shows a linear dependence on $\sqrt{eV/k_{B}T}$  when $eV\gg k_{B}T$. 


\section{Optical images of the SRO surface}
\label{appendix_Ru_inclusions}

\begin{figure}
\includegraphics[width=8cm]{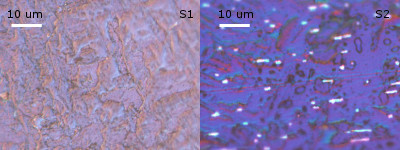}
\caption  {\small (Color online) Comparison of polarized optical microscope images of two samples: S1 (left), S2 (right). Ru inclusions are clearly seen in the right image for S2.  } 
\label{fig_micro_images}
\end{figure}

Optical images for SRO samples S1 and S2 are shown in Fig.~\ref{fig_micro_images} for comparison. Dense Ru inclusions of width about 1 $\mu$m and length a few $\mu$m are clearly seen in the micro image for S2, which is also harder to cleave than S1. This is consistent to the observation of Lichtenberg in Ref.~\onlinecite{Lichtenberg2002pssc} that SRO with Ru vacancies is much easier to cleave and the surface dead layer probably is also easier to pierce through. We note that although here the surface was polished by sandpaper to improve image quality, the Ru inclusions can easily be observed on the surface of S2 without any treatment.

\section{Reproducibility}
\label{appendix_reproducibility}

\begin{figure}
\includegraphics[width=9cm]{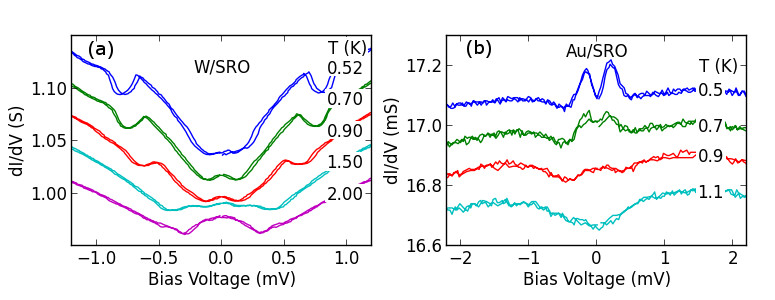}
\caption  {\small (Color online) (a) Temperature dependence of the PC spectra for a W/SRO PC on S1 but obtained in another run, showing similar gap features as in the main text. The resistance is only about 1 $\Omega$, but the gap around 0.15 mV, and the critical current effect are clearly demonstrated. (b) Temperature dependence of the PC spectra for a Au/SRO PC on S1, a split peak within $\pm$0.5 mV is observed, similar to that in  Ref.~[\onlinecite{Kashiwaya2011prl}].} 
\label{fig_other_PCS}
\end{figure}

PC spectra for more than 10 locations were measured in several runs. In each run a few locations are tried to search for SC-like features. With increasing force the tip eventually became blunted and bent, and small cracks can also develop on the surface of the SRO. A set of PC spectra similar to that in Fig.~\ref{fig_dVdI_4ohm} is shown in Fig.~\ref{fig_other_PCS}a, for a W/SRO PC on S1 but obtained in another run.  Besides W tip, Au tip (0.5 mm dia.) was also tried on S1 and the PC spectra are shown in Fig.~\ref{fig_other_PCS}b. For the Au/SRO PC, gap value around 0.5 mV is observed, the conductance enhancement is only about 1\%, and instead of the dome like conductance peak, a split peak is observed, similar to that was reported in Ref.~[\onlinecite{Kashiwaya2011prl}]. 

\begin{figure}
\includegraphics[width=9cm]{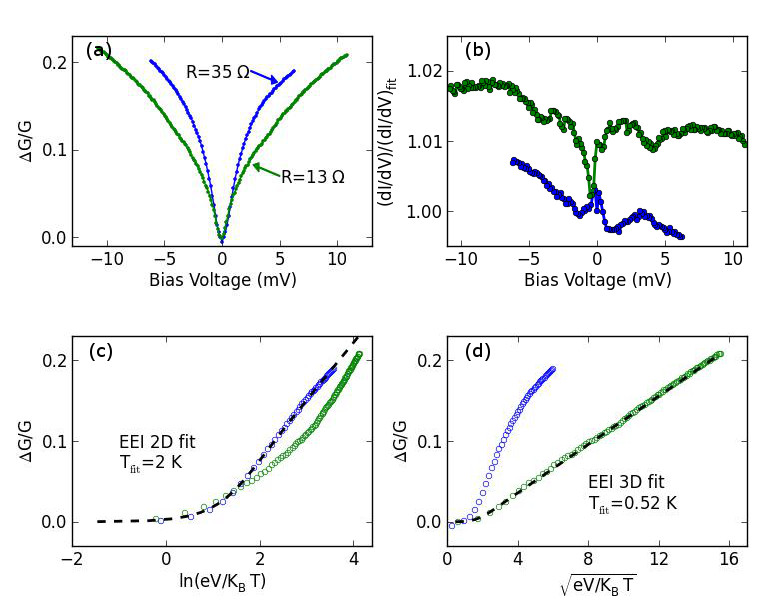}
\caption  {\small (Color online) (a) Bias dependence of the normalized conductance for point contact on S2 with resistance 35 (blue) and 13 $\Omega$  (green) at T=0. 52 K, and (b) the conductance after normalized by the background EEI fits (the green symbols are shifted up for clarity).  (c) EEI 2D fitting (black dash line) for the 13 $\Omega$ PC with  $T_{fit}$ = 2 K, slope 0.07. (d) EEI 3D fitting for the 35 $\Omega$ PC  with  $T_{fit}$ = 0.52 K, slope (3D) = 0.015. } 
\label{fig_two_kinds_ZBA}
\end{figure}

ZBA is less obvious for Au/SRO PCs. For W/SRO PCs, ZBA is frequently observed, which could be due to a thin oxide layer or a defective layer on the surface as observed in other PC measurements.~\cite{Sasaki2011prl,Peng2013prb} 
For those PC spectra showing clear ZBA, there are two typical types as shown in Fig.~\ref{fig_two_kinds_ZBA}.  One type is similar to that in Fig.~\ref{fig_dVdI_9ohm} with a logarithmic dependence consistent with 2D EEI, and SC feature sometimes coexists with ZBA; the other type has a $\sqrt{V}$ dependence which is consistent with 3D EEI, no SC feature is observed with this type of ZBA. 
For the 2D EEI type, e.g., for a 35 $\Omega$ PC on S2 as shown in Fig.~\ref{fig_two_kinds_ZBA}, the slope 0.07 is close to the slope 0.11 for S1 in Fig.~\ref{fig_dVdI_9ohm}, and 0.08 in Fig.~\ref{fig_dVdI_4ohm},  indicating similar 2D EEI is probed, though here $T_{fit}$ = 2 K is higher than the bath temperature about 0.52 K, which is probably the reason that SC feature is not observed. 

\begin{thebibliography}{57}%
\makeatletter
\providecommand \@ifxundefined [1]{%
 \@ifx{#1\undefined}
}%
\providecommand \@ifnum [1]{%
 \ifnum #1\expandafter \@firstoftwo
 \else \expandafter \@secondoftwo
 \fi
}%
\providecommand \@ifx [1]{%
 \ifx #1\expandafter \@firstoftwo
 \else \expandafter \@secondoftwo
 \fi
}%
\providecommand \natexlab [1]{#1}%
\providecommand \enquote  [1]{``#1''}%
\providecommand \bibnamefont  [1]{#1}%
\providecommand \bibfnamefont [1]{#1}%
\providecommand \citenamefont [1]{#1}%
\providecommand \href@noop [0]{\@secondoftwo}%
\providecommand \href [0]{\begingroup \@sanitize@url \@href}%
\providecommand \@href[1]{\@@startlink{#1}\@@href}%
\providecommand \@@href[1]{\endgroup#1\@@endlink}%
\providecommand \@sanitize@url [0]{\catcode `\\12\catcode `\$12\catcode
  `\&12\catcode `\#12\catcode `\^12\catcode `\_12\catcode `\%12\relax}%
\providecommand \@@startlink[1]{}%
\providecommand \@@endlink[0]{}%
\providecommand \url  [0]{\begingroup\@sanitize@url \@url }%
\providecommand \@url [1]{\endgroup\@href {#1}{\urlprefix }}%
\providecommand \urlprefix  [0]{URL }%
\providecommand \Eprint [0]{\href }%
\providecommand \doibase [0]{http://dx.doi.org/}%
\providecommand \selectlanguage [0]{\@gobble}%
\providecommand \bibinfo  [0]{\@secondoftwo}%
\providecommand \bibfield  [0]{\@secondoftwo}%
\providecommand \translation [1]{[#1]}%
\providecommand \BibitemOpen [0]{}%
\providecommand \bibitemStop [0]{}%
\providecommand \bibitemNoStop [0]{.\EOS\space}%
\providecommand \EOS [0]{\spacefactor3000\relax}%
\providecommand \BibitemShut  [1]{\csname bibitem#1\endcsname}%
\let\auto@bib@innerbib\@empty
\bibitem [{\citenamefont {{Maeno}}\ \emph {et~al.}(1994)\citenamefont
  {{Maeno}}, \citenamefont {{Hashimoto}}, \citenamefont {{Yoshida}},
  \citenamefont {{Nishizaki}}, \citenamefont {{Fujita}}, \citenamefont
  {{Bednorz}},\ and\ \citenamefont {{Lichtenberg}}}]{Maeno1994nature}%
  \BibitemOpen
  \bibfield  {author} {\bibinfo {author} {\bibfnamefont {Y.}~\bibnamefont
  {{Maeno}}}, \bibinfo {author} {\bibfnamefont {H.}~\bibnamefont
  {{Hashimoto}}}, \bibinfo {author} {\bibfnamefont {K.}~\bibnamefont
  {{Yoshida}}}, \bibinfo {author} {\bibfnamefont {S.}~\bibnamefont
  {{Nishizaki}}}, \bibinfo {author} {\bibfnamefont {T.}~\bibnamefont
  {{Fujita}}}, \bibinfo {author} {\bibfnamefont {J.~G.}\ \bibnamefont
  {{Bednorz}}}, \ and\ \bibinfo {author} {\bibfnamefont {F.}~\bibnamefont
  {{Lichtenberg}}},\ }\href {\doibase 10.1038/372532a0} {\bibfield  {journal}
  {\bibinfo  {journal} {Nature}\ }\textbf {\bibinfo {volume} {372}},\ \bibinfo
  {pages} {532} (\bibinfo {year} {1994})}\BibitemShut {NoStop}%
\bibitem [{\citenamefont {Mackenzie}\ and\ \citenamefont
  {Maeno}(2003)}]{Mackenzie2003rmp}%
  \BibitemOpen
  \bibfield  {author} {\bibinfo {author} {\bibfnamefont {A.~P.}\ \bibnamefont
  {Mackenzie}}\ and\ \bibinfo {author} {\bibfnamefont {Y.}~\bibnamefont
  {Maeno}},\ }\href {\doibase 10.1103/RevModPhys.75.657} {\bibfield  {journal}
  {\bibinfo  {journal} {Rev. Mod. Phys.}\ }\textbf {\bibinfo {volume} {75}},\
  \bibinfo {pages} {657} (\bibinfo {year} {2003})}\BibitemShut {NoStop}%
\bibitem [{\citenamefont {Maeno}\ \emph {et~al.}(2012)\citenamefont {Maeno},
  \citenamefont {Kittaka}, \citenamefont {Nomura}, \citenamefont {Yonezawa},\
  and\ \citenamefont {Ishida}}]{Maeno2012jpsj}%
  \BibitemOpen
  \bibfield  {author} {\bibinfo {author} {\bibfnamefont {Y.}~\bibnamefont
  {Maeno}}, \bibinfo {author} {\bibfnamefont {S.}~\bibnamefont {Kittaka}},
  \bibinfo {author} {\bibfnamefont {T.}~\bibnamefont {Nomura}}, \bibinfo
  {author} {\bibfnamefont {S.}~\bibnamefont {Yonezawa}}, \ and\ \bibinfo
  {author} {\bibfnamefont {K.}~\bibnamefont {Ishida}},\ }\href {\doibase
  10.1143/JPSJ.81.011009} {\bibfield  {journal} {\bibinfo  {journal} {Journal
  of the Physical Society of Japan}\ }\textbf {\bibinfo {volume} {81}},\
  \bibinfo {pages} {011009} (\bibinfo {year} {2012})},\ \Eprint
  {http://arxiv.org/abs/http://journals.jps.jp/doi/pdf/10.1143/JPSJ.81.011009}
  {http://journals.jps.jp/doi/pdf/10.1143/JPSJ.81.011009} \BibitemShut
  {NoStop}%
\bibitem [{\citenamefont {{Luke}}\ \emph {et~al.}(1998)\citenamefont {{Luke}},
  \citenamefont {{Fudamoto}}, \citenamefont {{Kojima}}, \citenamefont
  {{Larkin}}, \citenamefont {{Merrin}}, \citenamefont {{Nachumi}},
  \citenamefont {{Uemura}}, \citenamefont {{Maeno}}, \citenamefont {{Mao}},
  \citenamefont {{Mori}}, \citenamefont {{Nakamura}},\ and\ \citenamefont
  {{Sigrist}}}]{Luke1998nature}%
  \BibitemOpen
  \bibfield  {author} {\bibinfo {author} {\bibfnamefont {G.~M.}\ \bibnamefont
  {{Luke}}}, \bibinfo {author} {\bibfnamefont {Y.}~\bibnamefont {{Fudamoto}}},
  \bibinfo {author} {\bibfnamefont {K.~M.}\ \bibnamefont {{Kojima}}}, \bibinfo
  {author} {\bibfnamefont {M.~I.}\ \bibnamefont {{Larkin}}}, \bibinfo {author}
  {\bibfnamefont {J.}~\bibnamefont {{Merrin}}}, \bibinfo {author}
  {\bibfnamefont {B.}~\bibnamefont {{Nachumi}}}, \bibinfo {author}
  {\bibfnamefont {Y.~J.}\ \bibnamefont {{Uemura}}}, \bibinfo {author}
  {\bibfnamefont {Y.}~\bibnamefont {{Maeno}}}, \bibinfo {author} {\bibfnamefont
  {Z.~Q.}\ \bibnamefont {{Mao}}}, \bibinfo {author} {\bibfnamefont
  {Y.}~\bibnamefont {{Mori}}}, \bibinfo {author} {\bibfnamefont
  {H.}~\bibnamefont {{Nakamura}}}, \ and\ \bibinfo {author} {\bibfnamefont
  {M.}~\bibnamefont {{Sigrist}}},\ }\href {\doibase 10.1038/29038} {\bibfield
  {journal} {\bibinfo  {journal} {Nature}\ }\textbf {\bibinfo {volume} {394}},\
  \bibinfo {pages} {558} (\bibinfo {year} {1998})},\ \Eprint
  {http://arxiv.org/abs/cond-mat/9808159} {cond-mat/9808159} \BibitemShut
  {NoStop}%
\bibitem [{\citenamefont {Xia}\ \emph {et~al.}(2006)\citenamefont {Xia},
  \citenamefont {Maeno}, \citenamefont {Beyersdorf}, \citenamefont {Fejer},\
  and\ \citenamefont {Kapitulnik}}]{Xia2006prl}%
  \BibitemOpen
  \bibfield  {author} {\bibinfo {author} {\bibfnamefont {J.}~\bibnamefont
  {Xia}}, \bibinfo {author} {\bibfnamefont {Y.}~\bibnamefont {Maeno}}, \bibinfo
  {author} {\bibfnamefont {P.~T.}\ \bibnamefont {Beyersdorf}}, \bibinfo
  {author} {\bibfnamefont {M.~M.}\ \bibnamefont {Fejer}}, \ and\ \bibinfo
  {author} {\bibfnamefont {A.}~\bibnamefont {Kapitulnik}},\ }\href {\doibase
  10.1103/PhysRevLett.97.167002} {\bibfield  {journal} {\bibinfo  {journal}
  {Phys. Rev. Lett.}\ }\textbf {\bibinfo {volume} {97}},\ \bibinfo {pages}
  {167002} (\bibinfo {year} {2006})}\BibitemShut {NoStop}%
\bibitem [{\citenamefont {Kirtley}\ \emph {et~al.}(2007)\citenamefont
  {Kirtley}, \citenamefont {Kallin}, \citenamefont {Hicks}, \citenamefont
  {Kim}, \citenamefont {Liu}, \citenamefont {Moler}, \citenamefont {Maeno},\
  and\ \citenamefont {Nelson}}]{Kirtley2007prb}%
  \BibitemOpen
  \bibfield  {author} {\bibinfo {author} {\bibfnamefont {J.~R.}\ \bibnamefont
  {Kirtley}}, \bibinfo {author} {\bibfnamefont {C.}~\bibnamefont {Kallin}},
  \bibinfo {author} {\bibfnamefont {C.~W.}\ \bibnamefont {Hicks}}, \bibinfo
  {author} {\bibfnamefont {E.-A.}\ \bibnamefont {Kim}}, \bibinfo {author}
  {\bibfnamefont {Y.}~\bibnamefont {Liu}}, \bibinfo {author} {\bibfnamefont
  {K.~A.}\ \bibnamefont {Moler}}, \bibinfo {author} {\bibfnamefont
  {Y.}~\bibnamefont {Maeno}}, \ and\ \bibinfo {author} {\bibfnamefont {K.~D.}\
  \bibnamefont {Nelson}},\ }\href {\doibase 10.1103/PhysRevB.76.014526}
  {\bibfield  {journal} {\bibinfo  {journal} {Phys. Rev. B}\ }\textbf {\bibinfo
  {volume} {76}},\ \bibinfo {pages} {014526} (\bibinfo {year}
  {2007})}\BibitemShut {NoStop}%
\bibitem [{\citenamefont {Hicks}\ \emph {et~al.}(2010)\citenamefont {Hicks},
  \citenamefont {Kirtley}, \citenamefont {Lippman}, \citenamefont {Koshnick},
  \citenamefont {Huber}, \citenamefont {Maeno}, \citenamefont {Yuhasz},
  \citenamefont {Maple},\ and\ \citenamefont {Moler}}]{Hicks2010prb}%
  \BibitemOpen
  \bibfield  {author} {\bibinfo {author} {\bibfnamefont {C.~W.}\ \bibnamefont
  {Hicks}}, \bibinfo {author} {\bibfnamefont {J.~R.}\ \bibnamefont {Kirtley}},
  \bibinfo {author} {\bibfnamefont {T.~M.}\ \bibnamefont {Lippman}}, \bibinfo
  {author} {\bibfnamefont {N.~C.}\ \bibnamefont {Koshnick}}, \bibinfo {author}
  {\bibfnamefont {M.~E.}\ \bibnamefont {Huber}}, \bibinfo {author}
  {\bibfnamefont {Y.}~\bibnamefont {Maeno}}, \bibinfo {author} {\bibfnamefont
  {W.~M.}\ \bibnamefont {Yuhasz}}, \bibinfo {author} {\bibfnamefont {M.~B.}\
  \bibnamefont {Maple}}, \ and\ \bibinfo {author} {\bibfnamefont {K.~A.}\
  \bibnamefont {Moler}},\ }\href {\doibase 10.1103/PhysRevB.81.214501}
  {\bibfield  {journal} {\bibinfo  {journal} {Phys. Rev. B}\ }\textbf {\bibinfo
  {volume} {81}},\ \bibinfo {pages} {214501} (\bibinfo {year}
  {2010})}\BibitemShut {NoStop}%
\bibitem [{\citenamefont {Curran}\ \emph {et~al.}(2014)\citenamefont {Curran},
  \citenamefont {Bending}, \citenamefont {Desoky}, \citenamefont {Gibbs},
  \citenamefont {Lee},\ and\ \citenamefont {Mackenzie}}]{Curran2014prb}%
  \BibitemOpen
  \bibfield  {author} {\bibinfo {author} {\bibfnamefont {P.~J.}\ \bibnamefont
  {Curran}}, \bibinfo {author} {\bibfnamefont {S.~J.}\ \bibnamefont {Bending}},
  \bibinfo {author} {\bibfnamefont {W.~M.}\ \bibnamefont {Desoky}}, \bibinfo
  {author} {\bibfnamefont {A.~S.}\ \bibnamefont {Gibbs}}, \bibinfo {author}
  {\bibfnamefont {S.~L.}\ \bibnamefont {Lee}}, \ and\ \bibinfo {author}
  {\bibfnamefont {A.~P.}\ \bibnamefont {Mackenzie}},\ }\href {\doibase
  10.1103/PhysRevB.89.144504} {\bibfield  {journal} {\bibinfo  {journal} {Phys.
  Rev. B}\ }\textbf {\bibinfo {volume} {89}},\ \bibinfo {pages} {144504}
  (\bibinfo {year} {2014})}\BibitemShut {NoStop}%
\bibitem [{\citenamefont {Kashiwaya}\ \emph {et~al.}(2011)\citenamefont
  {Kashiwaya}, \citenamefont {Kashiwaya}, \citenamefont {Kambara},
  \citenamefont {Furuta}, \citenamefont {Yaguchi}, \citenamefont {Tanaka},\
  and\ \citenamefont {Maeno}}]{Kashiwaya2011prl}%
  \BibitemOpen
  \bibfield  {author} {\bibinfo {author} {\bibfnamefont {S.}~\bibnamefont
  {Kashiwaya}}, \bibinfo {author} {\bibfnamefont {H.}~\bibnamefont
  {Kashiwaya}}, \bibinfo {author} {\bibfnamefont {H.}~\bibnamefont {Kambara}},
  \bibinfo {author} {\bibfnamefont {T.}~\bibnamefont {Furuta}}, \bibinfo
  {author} {\bibfnamefont {H.}~\bibnamefont {Yaguchi}}, \bibinfo {author}
  {\bibfnamefont {Y.}~\bibnamefont {Tanaka}}, \ and\ \bibinfo {author}
  {\bibfnamefont {Y.}~\bibnamefont {Maeno}},\ }\href {\doibase
  10.1103/PhysRevLett.107.077003} {\bibfield  {journal} {\bibinfo  {journal}
  {Phys. Rev. Lett.}\ }\textbf {\bibinfo {volume} {107}},\ \bibinfo {pages}
  {077003} (\bibinfo {year} {2011})}\BibitemShut {NoStop}%
\bibitem [{\citenamefont {Kashiwaya}\ \emph {et~al.}(2014)\citenamefont
  {Kashiwaya}, \citenamefont {Kashiwaya}, \citenamefont {Saitoh}, \citenamefont
  {Mawatari},\ and\ \citenamefont {Tanaka}}]{Kashiwaya2014pe}%
  \BibitemOpen
  \bibfield  {author} {\bibinfo {author} {\bibfnamefont {S.}~\bibnamefont
  {Kashiwaya}}, \bibinfo {author} {\bibfnamefont {H.}~\bibnamefont
  {Kashiwaya}}, \bibinfo {author} {\bibfnamefont {K.}~\bibnamefont {Saitoh}},
  \bibinfo {author} {\bibfnamefont {Y.}~\bibnamefont {Mawatari}}, \ and\
  \bibinfo {author} {\bibfnamefont {Y.}~\bibnamefont {Tanaka}},\ }\href
  {\doibase http://dx.doi.org/10.1016/j.physe.2013.07.016} {\bibfield
  {journal} {\bibinfo  {journal} {Physica E: Low-dimensional Systems and
  Nanostructures}\ }\textbf {\bibinfo {volume} {55}},\ \bibinfo {pages} {25 }
  (\bibinfo {year} {2014})},\ \bibinfo {note} {topological Objects}\BibitemShut
  {NoStop}%
\bibitem [{\citenamefont {Laube}\ \emph {et~al.}(2000)\citenamefont {Laube},
  \citenamefont {Goll}, \citenamefont {L\"ohneysen}, \citenamefont
  {Fogelstr\"om},\ and\ \citenamefont {Lichtenberg}}]{Laube2000prl}%
  \BibitemOpen
  \bibfield  {author} {\bibinfo {author} {\bibfnamefont {F.}~\bibnamefont
  {Laube}}, \bibinfo {author} {\bibfnamefont {G.}~\bibnamefont {Goll}},
  \bibinfo {author} {\bibfnamefont {H.~v.}\ \bibnamefont {L\"ohneysen}},
  \bibinfo {author} {\bibfnamefont {M.}~\bibnamefont {Fogelstr\"om}}, \ and\
  \bibinfo {author} {\bibfnamefont {F.}~\bibnamefont {Lichtenberg}},\ }\href
  {\doibase 10.1103/PhysRevLett.84.1595} {\bibfield  {journal} {\bibinfo
  {journal} {Phys. Rev. Lett.}\ }\textbf {\bibinfo {volume} {84}},\ \bibinfo
  {pages} {1595} (\bibinfo {year} {2000})}\BibitemShut {NoStop}%
\bibitem [{\citenamefont {Upward}\ \emph {et~al.}(2002)\citenamefont {Upward},
  \citenamefont {Kouwenhoven}, \citenamefont {Morpurgo}, \citenamefont
  {Kikugawa}, \citenamefont {Mao},\ and\ \citenamefont
  {Maeno}}]{Upward2002prb}%
  \BibitemOpen
  \bibfield  {author} {\bibinfo {author} {\bibfnamefont {M.~D.}\ \bibnamefont
  {Upward}}, \bibinfo {author} {\bibfnamefont {L.~P.}\ \bibnamefont
  {Kouwenhoven}}, \bibinfo {author} {\bibfnamefont {A.~F.}\ \bibnamefont
  {Morpurgo}}, \bibinfo {author} {\bibfnamefont {N.}~\bibnamefont {Kikugawa}},
  \bibinfo {author} {\bibfnamefont {Z.~Q.}\ \bibnamefont {Mao}}, \ and\
  \bibinfo {author} {\bibfnamefont {Y.}~\bibnamefont {Maeno}},\ }\href
  {\doibase 10.1103/PhysRevB.65.220512} {\bibfield  {journal} {\bibinfo
  {journal} {Phys. Rev. B}\ }\textbf {\bibinfo {volume} {65}},\ \bibinfo
  {pages} {220512} (\bibinfo {year} {2002})}\BibitemShut {NoStop}%
\bibitem [{\citenamefont {Firmo}\ \emph {et~al.}(2013)\citenamefont {Firmo},
  \citenamefont {Lederer}, \citenamefont {Lupien}, \citenamefont {Mackenzie},
  \citenamefont {Davis},\ and\ \citenamefont {Kivelson}}]{Firmo2013prb}%
  \BibitemOpen
  \bibfield  {author} {\bibinfo {author} {\bibfnamefont {I.~A.}\ \bibnamefont
  {Firmo}}, \bibinfo {author} {\bibfnamefont {S.}~\bibnamefont {Lederer}},
  \bibinfo {author} {\bibfnamefont {C.}~\bibnamefont {Lupien}}, \bibinfo
  {author} {\bibfnamefont {A.~P.}\ \bibnamefont {Mackenzie}}, \bibinfo {author}
  {\bibfnamefont {J.~C.}\ \bibnamefont {Davis}}, \ and\ \bibinfo {author}
  {\bibfnamefont {S.~A.}\ \bibnamefont {Kivelson}},\ }\href {\doibase
  10.1103/PhysRevB.88.134521} {\bibfield  {journal} {\bibinfo  {journal} {Phys.
  Rev. B}\ }\textbf {\bibinfo {volume} {88}},\ \bibinfo {pages} {134521}
  (\bibinfo {year} {2013})}\BibitemShut {NoStop}%
\bibitem [{\citenamefont {Matzdorf}\ \emph {et~al.}(2000)\citenamefont
  {Matzdorf}, \citenamefont {Fang}, \citenamefont {Ismail}, \citenamefont
  {Zhang}, \citenamefont {Kimura}, \citenamefont {Tokura}, \citenamefont
  {Terakura},\ and\ \citenamefont {Plummer}}]{Matzdorf2000science}%
  \BibitemOpen
  \bibfield  {author} {\bibinfo {author} {\bibfnamefont {R.}~\bibnamefont
  {Matzdorf}}, \bibinfo {author} {\bibfnamefont {Z.}~\bibnamefont {Fang}},
  \bibinfo {author} {\bibnamefont {Ismail}}, \bibinfo {author} {\bibfnamefont
  {J.}~\bibnamefont {Zhang}}, \bibinfo {author} {\bibfnamefont
  {T.}~\bibnamefont {Kimura}}, \bibinfo {author} {\bibfnamefont
  {Y.}~\bibnamefont {Tokura}}, \bibinfo {author} {\bibfnamefont
  {K.}~\bibnamefont {Terakura}}, \ and\ \bibinfo {author} {\bibfnamefont
  {E.~W.}\ \bibnamefont {Plummer}},\ }\href {\doibase
  10.1126/science.289.5480.746} {\bibfield  {journal} {\bibinfo  {journal}
  {Science}\ }\textbf {\bibinfo {volume} {289}},\ \bibinfo {pages} {746}
  (\bibinfo {year} {2000})},\ \Eprint
  {http://arxiv.org/abs/http://www.sciencemag.org/content/289/5480/746.full.pdf}
  {http://www.sciencemag.org/content/289/5480/746.full.pdf} \BibitemShut
  {NoStop}%
\bibitem [{\citenamefont {Lederer}\ \emph {et~al.}(2014)\citenamefont
  {Lederer}, \citenamefont {Huang}, \citenamefont {Taylor}, \citenamefont
  {Raghu},\ and\ \citenamefont {Kallin}}]{Lederer2014prb}%
  \BibitemOpen
  \bibfield  {author} {\bibinfo {author} {\bibfnamefont {S.}~\bibnamefont
  {Lederer}}, \bibinfo {author} {\bibfnamefont {W.}~\bibnamefont {Huang}},
  \bibinfo {author} {\bibfnamefont {E.}~\bibnamefont {Taylor}}, \bibinfo
  {author} {\bibfnamefont {S.}~\bibnamefont {Raghu}}, \ and\ \bibinfo {author}
  {\bibfnamefont {C.}~\bibnamefont {Kallin}},\ }\href {\doibase
  10.1103/PhysRevB.90.134521} {\bibfield  {journal} {\bibinfo  {journal} {Phys.
  Rev. B}\ }\textbf {\bibinfo {volume} {90}},\ \bibinfo {pages} {134521}
  (\bibinfo {year} {2014})}\BibitemShut {NoStop}%
\bibitem [{\citenamefont {{Gonnelli}}\ \emph {et~al.}(2002)\citenamefont
  {{Gonnelli}}, \citenamefont {{Calzolari}}, \citenamefont {{Daghero}},
  \citenamefont {{Ummarino}}, \citenamefont {{Stepanov}}, \citenamefont
  {{Fino}}, \citenamefont {{Giunchi}}, \citenamefont {{Ceresara}},\ and\
  \citenamefont {{Ripamonti}}}]{Gonnelli2002jpcs}%
  \BibitemOpen
  \bibfield  {author} {\bibinfo {author} {\bibfnamefont {R.~S.}\ \bibnamefont
  {{Gonnelli}}}, \bibinfo {author} {\bibfnamefont {A.}~\bibnamefont
  {{Calzolari}}}, \bibinfo {author} {\bibfnamefont {D.}~\bibnamefont
  {{Daghero}}}, \bibinfo {author} {\bibfnamefont {G.~A.}\ \bibnamefont
  {{Ummarino}}}, \bibinfo {author} {\bibfnamefont {V.~A.}\ \bibnamefont
  {{Stepanov}}}, \bibinfo {author} {\bibfnamefont {P.}~\bibnamefont {{Fino}}},
  \bibinfo {author} {\bibfnamefont {G.}~\bibnamefont {{Giunchi}}}, \bibinfo
  {author} {\bibfnamefont {S.}~\bibnamefont {{Ceresara}}}, \ and\ \bibinfo
  {author} {\bibfnamefont {G.}~\bibnamefont {{Ripamonti}}},\ }\href {\doibase
  10.1016/S0022-3697(02)00229-9} {\bibfield  {journal} {\bibinfo  {journal}
  {Journal of Physics and Chemistry of Solids}\ }\textbf {\bibinfo {volume}
  {63}},\ \bibinfo {pages} {2319} (\bibinfo {year} {2002})},\ \Eprint
  {http://arxiv.org/abs/cond-mat/0107239} {cond-mat/0107239} \BibitemShut
  {NoStop}%
\bibitem [{\citenamefont {Gloos}\ \emph {et~al.}(1996)\citenamefont {Gloos},
  \citenamefont {Anders}, \citenamefont {Buschinger}, \citenamefont {Geibel},
  \citenamefont {Heuser}, \citenamefont {J\"{a}hrling}, \citenamefont {Kim},
  \citenamefont {Klemens}, \citenamefont {M\"{u}ller-Reisener}, \citenamefont
  {Schank},\ and\ \citenamefont {Stewart}}]{Gloos1996jltp_scaling}%
  \BibitemOpen
  \bibfield  {author} {\bibinfo {author} {\bibfnamefont {K.}~\bibnamefont
  {Gloos}}, \bibinfo {author} {\bibfnamefont {F.}~\bibnamefont {Anders}},
  \bibinfo {author} {\bibfnamefont {B.}~\bibnamefont {Buschinger}}, \bibinfo
  {author} {\bibfnamefont {C.}~\bibnamefont {Geibel}}, \bibinfo {author}
  {\bibfnamefont {K.}~\bibnamefont {Heuser}}, \bibinfo {author} {\bibfnamefont
  {F.}~\bibnamefont {J\"{a}hrling}}, \bibinfo {author} {\bibfnamefont
  {J.}~\bibnamefont {Kim}}, \bibinfo {author} {\bibfnamefont {R.}~\bibnamefont
  {Klemens}}, \bibinfo {author} {\bibfnamefont {R.}~\bibnamefont
  {M\"{u}ller-Reisener}}, \bibinfo {author} {\bibfnamefont {C.}~\bibnamefont
  {Schank}}, \ and\ \bibinfo {author} {\bibfnamefont {G.}~\bibnamefont
  {Stewart}},\ }\href {\doibase 10.1007/BF00754627} {\bibfield  {journal}
  {\bibinfo  {journal} {Journal of Low Temperature Physics}\ }\textbf {\bibinfo
  {volume} {105}},\ \bibinfo {pages} {37} (\bibinfo {year} {1996})}\BibitemShut
  {NoStop}%
\bibitem [{\citenamefont {Daghero}\ and\ \citenamefont
  {Gonnelli}(2010)}]{Daghero2010sst}%
  \BibitemOpen
  \bibfield  {author} {\bibinfo {author} {\bibfnamefont {D.}~\bibnamefont
  {Daghero}}\ and\ \bibinfo {author} {\bibfnamefont {R.~S.}\ \bibnamefont
  {Gonnelli}},\ }\href {http://stacks.iop.org/0953-2048/23/i=4/a=043001}
  {\bibfield  {journal} {\bibinfo  {journal} {Superconductor Science and
  Technology}\ }\textbf {\bibinfo {volume} {23}},\ \bibinfo {pages} {043001}
  (\bibinfo {year} {2010})}\BibitemShut {NoStop}%
\bibitem [{\citenamefont {Gloos}\ \emph {et~al.}(1995)\citenamefont {Gloos},
  \citenamefont {Martin}, \citenamefont {Schank}, \citenamefont {Geibel},\ and\
  \citenamefont {Steglich}}]{Gloos1995pb}%
  \BibitemOpen
  \bibfield  {author} {\bibinfo {author} {\bibfnamefont {K.}~\bibnamefont
  {Gloos}}, \bibinfo {author} {\bibfnamefont {F.}~\bibnamefont {Martin}},
  \bibinfo {author} {\bibfnamefont {C.}~\bibnamefont {Schank}}, \bibinfo
  {author} {\bibfnamefont {C.}~\bibnamefont {Geibel}}, \ and\ \bibinfo {author}
  {\bibfnamefont {F.}~\bibnamefont {Steglich}},\ }\href {\doibase
  dx.doi.org/10.1016/0921-4526(94)00433-V} {\bibfield  {journal} {\bibinfo
  {journal} {Physica B: Condensed Matter}\ }\textbf {\bibinfo {volume}
  {206-207}},\ \bibinfo {pages} {279 } (\bibinfo {year} {1995})},\ \bibinfo
  {note} {proceedings of the International Conference on Strongly Correlated
  Electron Systems}\BibitemShut {NoStop}%
\bibitem [{\citenamefont {Miyoshi}\ \emph {et~al.}(2005)\citenamefont
  {Miyoshi}, \citenamefont {Bugoslavsky},\ and\ \citenamefont
  {Cohen}}]{Miyoshi2005prb}%
  \BibitemOpen
  \bibfield  {author} {\bibinfo {author} {\bibfnamefont {Y.}~\bibnamefont
  {Miyoshi}}, \bibinfo {author} {\bibfnamefont {Y.}~\bibnamefont
  {Bugoslavsky}}, \ and\ \bibinfo {author} {\bibfnamefont {L.~F.}\ \bibnamefont
  {Cohen}},\ }\href {\doibase 10.1103/PhysRevB.72.012502} {\bibfield  {journal}
  {\bibinfo  {journal} {Phys. Rev. B}\ }\textbf {\bibinfo {volume} {72}},\
  \bibinfo {pages} {012502} (\bibinfo {year} {2005})}\BibitemShut {NoStop}%
\bibitem [{\citenamefont {Kittaka}\ \emph {et~al.}(2010)\citenamefont
  {Kittaka}, \citenamefont {Taniguchi}, \citenamefont {Yonezawa}, \citenamefont
  {Yaguchi},\ and\ \citenamefont {Maeno}}]{Kittaka2010prb}%
  \BibitemOpen
  \bibfield  {author} {\bibinfo {author} {\bibfnamefont {S.}~\bibnamefont
  {Kittaka}}, \bibinfo {author} {\bibfnamefont {H.}~\bibnamefont {Taniguchi}},
  \bibinfo {author} {\bibfnamefont {S.}~\bibnamefont {Yonezawa}}, \bibinfo
  {author} {\bibfnamefont {H.}~\bibnamefont {Yaguchi}}, \ and\ \bibinfo
  {author} {\bibfnamefont {Y.}~\bibnamefont {Maeno}},\ }\href {\doibase
  10.1103/PhysRevB.81.180510} {\bibfield  {journal} {\bibinfo  {journal} {Phys.
  Rev. B}\ }\textbf {\bibinfo {volume} {81}},\ \bibinfo {pages} {180510}
  (\bibinfo {year} {2010})}\BibitemShut {NoStop}%
\bibitem [{\citenamefont {Kittaka}\ \emph {et~al.}(2009)\citenamefont
  {Kittaka}, \citenamefont {Yaguchi},\ and\ \citenamefont
  {Maeno}}]{Kittaka2009jpsj_b}%
  \BibitemOpen
  \bibfield  {author} {\bibinfo {author} {\bibfnamefont {S.}~\bibnamefont
  {Kittaka}}, \bibinfo {author} {\bibfnamefont {H.}~\bibnamefont {Yaguchi}}, \
  and\ \bibinfo {author} {\bibfnamefont {Y.}~\bibnamefont {Maeno}},\ }\href
  {\doibase 10.1143/JPSJ.78.103705} {\bibfield  {journal} {\bibinfo  {journal}
  {Journal of the Physical Society of Japan}\ }\textbf {\bibinfo {volume}
  {78}},\ \bibinfo {pages} {103705} (\bibinfo {year} {2009})},\ \Eprint
  {http://arxiv.org/abs/http://journals.jps.jp/doi/pdf/10.1143/JPSJ.78.103705}
  {http://journals.jps.jp/doi/pdf/10.1143/JPSJ.78.103705} \BibitemShut
  {NoStop}%
\bibitem [{\citenamefont {Hicks}\ \emph {et~al.}(2014)\citenamefont {Hicks},
  \citenamefont {Brodsky}, \citenamefont {Yelland}, \citenamefont {Gibbs},
  \citenamefont {Bruin}, \citenamefont {Barber}, \citenamefont {Edkins},
  \citenamefont {Nishimura}, \citenamefont {Yonezawa}, \citenamefont {Maeno},\
  and\ \citenamefont {Mackenzie}}]{Hicks2014science}%
  \BibitemOpen
  \bibfield  {author} {\bibinfo {author} {\bibfnamefont {C.~W.}\ \bibnamefont
  {Hicks}}, \bibinfo {author} {\bibfnamefont {D.~O.}\ \bibnamefont {Brodsky}},
  \bibinfo {author} {\bibfnamefont {E.~A.}\ \bibnamefont {Yelland}}, \bibinfo
  {author} {\bibfnamefont {A.~S.}\ \bibnamefont {Gibbs}}, \bibinfo {author}
  {\bibfnamefont {J.~A.~N.}\ \bibnamefont {Bruin}}, \bibinfo {author}
  {\bibfnamefont {M.~E.}\ \bibnamefont {Barber}}, \bibinfo {author}
  {\bibfnamefont {S.~D.}\ \bibnamefont {Edkins}}, \bibinfo {author}
  {\bibfnamefont {K.}~\bibnamefont {Nishimura}}, \bibinfo {author}
  {\bibfnamefont {S.}~\bibnamefont {Yonezawa}}, \bibinfo {author}
  {\bibfnamefont {Y.}~\bibnamefont {Maeno}}, \ and\ \bibinfo {author}
  {\bibfnamefont {A.~P.}\ \bibnamefont {Mackenzie}},\ }\href {\doibase
  10.1126/science.1248292} {\bibfield  {journal} {\bibinfo  {journal}
  {Science}\ }\textbf {\bibinfo {volume} {344}},\ \bibinfo {pages} {283}
  (\bibinfo {year} {2014})},\ \Eprint
  {http://arxiv.org/abs/http://www.sciencemag.org/content/344/6181/283.full.pdf}
  {http://www.sciencemag.org/content/344/6181/283.full.pdf} \BibitemShut
  {NoStop}%
\bibitem [{\citenamefont {Mao}\ \emph {et~al.}(2000)\citenamefont {Mao},
  \citenamefont {Maenoab},\ and\ \citenamefont {Fukazawa}}]{Mao2000mrb}%
  \BibitemOpen
  \bibfield  {author} {\bibinfo {author} {\bibfnamefont {Z.}~\bibnamefont
  {Mao}}, \bibinfo {author} {\bibfnamefont {Y.}~\bibnamefont {Maenoab}}, \ and\
  \bibinfo {author} {\bibfnamefont {H.}~\bibnamefont {Fukazawa}},\ }\href
  {\doibase 10.1016/S0025-5408(00)00378-0} {\bibfield  {journal} {\bibinfo
  {journal} {Materials Research Bulletin}\ }\textbf {\bibinfo {volume} {35}},\
  \bibinfo {pages} {1813 } (\bibinfo {year} {2000})}\BibitemShut {NoStop}%
\bibitem [{\citenamefont {Taniguchi}\ \emph {et~al.}(2012)\citenamefont
  {Taniguchi}, \citenamefont {Kittaka}, \citenamefont {Yonezawa}, \citenamefont
  {Yaguchi},\ and\ \citenamefont {Maeno}}]{Taniguchi2012jpcs}%
  \BibitemOpen
  \bibfield  {author} {\bibinfo {author} {\bibfnamefont {H.}~\bibnamefont
  {Taniguchi}}, \bibinfo {author} {\bibfnamefont {S.}~\bibnamefont {Kittaka}},
  \bibinfo {author} {\bibfnamefont {S.}~\bibnamefont {Yonezawa}}, \bibinfo
  {author} {\bibfnamefont {H.}~\bibnamefont {Yaguchi}}, \ and\ \bibinfo
  {author} {\bibfnamefont {Y.}~\bibnamefont {Maeno}},\ }\href
  {http://stacks.iop.org/1742-6596/391/i=1/a=012108} {\bibfield  {journal}
  {\bibinfo  {journal} {Journal of Physics: Conference Series}\ }\textbf
  {\bibinfo {volume} {391}},\ \bibinfo {pages} {012108} (\bibinfo {year}
  {2012})}\BibitemShut {NoStop}%
\bibitem [{\citenamefont {Ying}\ \emph {et~al.}(2009)\citenamefont {Ying},
  \citenamefont {Xin}, \citenamefont {Clouser}, \citenamefont {Hao},
  \citenamefont {Staley}, \citenamefont {Myers}, \citenamefont {Allard},
  \citenamefont {Fobes}, \citenamefont {Liu}, \citenamefont {Mao},\ and\
  \citenamefont {Liu}}]{Ying2009prl}%
  \BibitemOpen
  \bibfield  {author} {\bibinfo {author} {\bibfnamefont {Y.~A.}\ \bibnamefont
  {Ying}}, \bibinfo {author} {\bibfnamefont {Y.}~\bibnamefont {Xin}}, \bibinfo
  {author} {\bibfnamefont {B.~W.}\ \bibnamefont {Clouser}}, \bibinfo {author}
  {\bibfnamefont {E.}~\bibnamefont {Hao}}, \bibinfo {author} {\bibfnamefont
  {N.~E.}\ \bibnamefont {Staley}}, \bibinfo {author} {\bibfnamefont {R.~J.}\
  \bibnamefont {Myers}}, \bibinfo {author} {\bibfnamefont {L.~F.}\ \bibnamefont
  {Allard}}, \bibinfo {author} {\bibfnamefont {D.}~\bibnamefont {Fobes}},
  \bibinfo {author} {\bibfnamefont {T.}~\bibnamefont {Liu}}, \bibinfo {author}
  {\bibfnamefont {Z.~Q.}\ \bibnamefont {Mao}}, \ and\ \bibinfo {author}
  {\bibfnamefont {Y.}~\bibnamefont {Liu}},\ }\href {\doibase
  10.1103/PhysRevLett.103.247004} {\bibfield  {journal} {\bibinfo  {journal}
  {Phys. Rev. Lett.}\ }\textbf {\bibinfo {volume} {103}},\ \bibinfo {pages}
  {247004} (\bibinfo {year} {2009})}\BibitemShut {NoStop}%
\bibitem [{\citenamefont {{Ying}}\ \emph {et~al.}(2013)\citenamefont {{Ying}},
  \citenamefont {{Staley}}, \citenamefont {{Xin}}, \citenamefont {{Sun}},
  \citenamefont {{Cai}}, \citenamefont {{Fobes}}, \citenamefont {{Liu}},
  \citenamefont {{Mao}},\ and\ \citenamefont {{Liu}}}]{Ying2013ncomms}%
  \BibitemOpen
  \bibfield  {author} {\bibinfo {author} {\bibfnamefont {Y.~A.}\ \bibnamefont
  {{Ying}}}, \bibinfo {author} {\bibfnamefont {N.~E.}\ \bibnamefont
  {{Staley}}}, \bibinfo {author} {\bibfnamefont {Y.}~\bibnamefont {{Xin}}},
  \bibinfo {author} {\bibfnamefont {K.}~\bibnamefont {{Sun}}}, \bibinfo
  {author} {\bibfnamefont {X.}~\bibnamefont {{Cai}}}, \bibinfo {author}
  {\bibfnamefont {D.}~\bibnamefont {{Fobes}}}, \bibinfo {author} {\bibfnamefont
  {T.~J.}\ \bibnamefont {{Liu}}}, \bibinfo {author} {\bibfnamefont {Z.~Q.}\
  \bibnamefont {{Mao}}}, \ and\ \bibinfo {author} {\bibfnamefont
  {Y.}~\bibnamefont {{Liu}}},\ }\href {\doibase 10.1038/ncomms3596} {\bibfield
  {journal} {\bibinfo  {journal} {Nature Communications}\ }\textbf {\bibinfo
  {volume} {4}},\ \bibinfo {eid} {2596} (\bibinfo {year} {2013}),\
  10.1038/ncomms3596}\BibitemShut {NoStop}%
\bibitem [{\citenamefont {Dikin}\ \emph {et~al.}(2011)\citenamefont {Dikin},
  \citenamefont {Mehta}, \citenamefont {Bark}, \citenamefont {Folkman},
  \citenamefont {Eom},\ and\ \citenamefont {Chandrasekhar}}]{Dikin2011prl}%
  \BibitemOpen
  \bibfield  {author} {\bibinfo {author} {\bibfnamefont {D.~A.}\ \bibnamefont
  {Dikin}}, \bibinfo {author} {\bibfnamefont {M.}~\bibnamefont {Mehta}},
  \bibinfo {author} {\bibfnamefont {C.~W.}\ \bibnamefont {Bark}}, \bibinfo
  {author} {\bibfnamefont {C.~M.}\ \bibnamefont {Folkman}}, \bibinfo {author}
  {\bibfnamefont {C.~B.}\ \bibnamefont {Eom}}, \ and\ \bibinfo {author}
  {\bibfnamefont {V.}~\bibnamefont {Chandrasekhar}},\ }\href {\doibase
  10.1103/PhysRevLett.107.056802} {\bibfield  {journal} {\bibinfo  {journal}
  {Phys. Rev. Lett.}\ }\textbf {\bibinfo {volume} {107}},\ \bibinfo {pages}
  {056802} (\bibinfo {year} {2011})}\BibitemShut {NoStop}%
\bibitem [{\citenamefont {Deguchi}\ \emph {et~al.}(2004)\citenamefont
  {Deguchi}, \citenamefont {Q.~Mao},\ and\ \citenamefont
  {Maeno}}]{Deguchi2004jpsj}%
  \BibitemOpen
  \bibfield  {author} {\bibinfo {author} {\bibfnamefont {K.}~\bibnamefont
  {Deguchi}}, \bibinfo {author} {\bibfnamefont {Z.}~\bibnamefont {Q.~Mao}}, \
  and\ \bibinfo {author} {\bibfnamefont {Y.}~\bibnamefont {Maeno}},\ }\href
  {\doibase 10.1143/JPSJ.73.1313} {\bibfield  {journal} {\bibinfo  {journal}
  {Journal of the Physical Society of Japan}\ }\textbf {\bibinfo {volume}
  {73}},\ \bibinfo {pages} {1313} (\bibinfo {year} {2004})}\BibitemShut
  {NoStop}%
\bibitem [{\citenamefont {Dumont}\ and\ \citenamefont
  {Mota}(2002)}]{Dumont2002prb}%
  \BibitemOpen
  \bibfield  {author} {\bibinfo {author} {\bibfnamefont {E.}~\bibnamefont
  {Dumont}}\ and\ \bibinfo {author} {\bibfnamefont {A.~C.}\ \bibnamefont
  {Mota}},\ }\href {\doibase 10.1103/PhysRevB.65.144519} {\bibfield  {journal}
  {\bibinfo  {journal} {Phys. Rev. B}\ }\textbf {\bibinfo {volume} {65}},\
  \bibinfo {pages} {144519} (\bibinfo {year} {2002})}\BibitemShut {NoStop}%
\bibitem [{\citenamefont {Kambara}\ \emph {et~al.}(2008)\citenamefont
  {Kambara}, \citenamefont {Kashiwaya}, \citenamefont {Yaguchi}, \citenamefont
  {Asano}, \citenamefont {Tanaka},\ and\ \citenamefont
  {Maeno}}]{Kambara2008prl}%
  \BibitemOpen
  \bibfield  {author} {\bibinfo {author} {\bibfnamefont {H.}~\bibnamefont
  {Kambara}}, \bibinfo {author} {\bibfnamefont {S.}~\bibnamefont {Kashiwaya}},
  \bibinfo {author} {\bibfnamefont {H.}~\bibnamefont {Yaguchi}}, \bibinfo
  {author} {\bibfnamefont {Y.}~\bibnamefont {Asano}}, \bibinfo {author}
  {\bibfnamefont {Y.}~\bibnamefont {Tanaka}}, \ and\ \bibinfo {author}
  {\bibfnamefont {Y.}~\bibnamefont {Maeno}},\ }\href {\doibase
  10.1103/PhysRevLett.101.267003} {\bibfield  {journal} {\bibinfo  {journal}
  {Phys. Rev. Lett.}\ }\textbf {\bibinfo {volume} {101}},\ \bibinfo {pages}
  {267003} (\bibinfo {year} {2008})}\BibitemShut {NoStop}%
\bibitem [{\citenamefont {Ikeda}\ \emph {et~al.}(2000)\citenamefont {Ikeda},
  \citenamefont {Maeno}, \citenamefont {Nakatsuji}, \citenamefont {Kosaka},\
  and\ \citenamefont {Uwatoko}}]{Ikeda2000prb}%
  \BibitemOpen
  \bibfield  {author} {\bibinfo {author} {\bibfnamefont {S.-I.}\ \bibnamefont
  {Ikeda}}, \bibinfo {author} {\bibfnamefont {Y.}~\bibnamefont {Maeno}},
  \bibinfo {author} {\bibfnamefont {S.}~\bibnamefont {Nakatsuji}}, \bibinfo
  {author} {\bibfnamefont {M.}~\bibnamefont {Kosaka}}, \ and\ \bibinfo {author}
  {\bibfnamefont {Y.}~\bibnamefont {Uwatoko}},\ }\href {\doibase
  10.1103/PhysRevB.62.R6089} {\bibfield  {journal} {\bibinfo  {journal} {Phys.
  Rev. B}\ }\textbf {\bibinfo {volume} {62}},\ \bibinfo {pages} {R6089}
  (\bibinfo {year} {2000})}\BibitemShut {NoStop}%
\bibitem [{\citenamefont {{Ikeda}}\ \emph {et~al.}(2004)\citenamefont
  {{Ikeda}}, \citenamefont {{Koiwai}}, \citenamefont {{Yoshida}}, \citenamefont
  {{Shirakawa}}, \citenamefont {{Hara}}, \citenamefont {{Kosaka}},\ and\
  \citenamefont {{Uwatoko}}}]{Ikeda2004jmmm}%
  \BibitemOpen
  \bibfield  {author} {\bibinfo {author} {\bibfnamefont {S.~I.}\ \bibnamefont
  {{Ikeda}}}, \bibinfo {author} {\bibfnamefont {S.}~\bibnamefont {{Koiwai}}},
  \bibinfo {author} {\bibfnamefont {Y.}~\bibnamefont {{Yoshida}}}, \bibinfo
  {author} {\bibfnamefont {N.}~\bibnamefont {{Shirakawa}}}, \bibinfo {author}
  {\bibfnamefont {S.}~\bibnamefont {{Hara}}}, \bibinfo {author} {\bibfnamefont
  {M.}~\bibnamefont {{Kosaka}}}, \ and\ \bibinfo {author} {\bibfnamefont
  {Y.}~\bibnamefont {{Uwatoko}}},\ }\href {\doibase
  10.1016/j.jmmm.2003.12.1027} {\bibfield  {journal} {\bibinfo  {journal}
  {Journal of Magnetism and Magnetic Materials}\ }\textbf {\bibinfo {volume}
  {272}} (\bibinfo {year} {2004}),\ 10.1016/j.jmmm.2003.12.1027}\BibitemShut
  {NoStop}%
\bibitem [{\citenamefont {{Carlo}}\ \emph {et~al.}(2012)\citenamefont
  {{Carlo}}, \citenamefont {{Goko}}, \citenamefont {{Gat-Malureanu}},
  \citenamefont {{Russo}}, \citenamefont {{Savici}}, \citenamefont {{Aczel}},
  \citenamefont {{MacDougall}}, \citenamefont {{Rodriguez}}, \citenamefont
  {{Williams}}, \citenamefont {{Luke}}, \citenamefont {{Wiebe}}, \citenamefont
  {{Yoshida}}, \citenamefont {{Nakatsuji}}, \citenamefont {{Maeno}},
  \citenamefont {{Taniguchi}},\ and\ \citenamefont {{Uemura}}}]{Carlo2012nmat}%
  \BibitemOpen
  \bibfield  {author} {\bibinfo {author} {\bibfnamefont {J.~P.}\ \bibnamefont
  {{Carlo}}}, \bibinfo {author} {\bibfnamefont {T.}~\bibnamefont {{Goko}}},
  \bibinfo {author} {\bibfnamefont {I.~M.}\ \bibnamefont {{Gat-Malureanu}}},
  \bibinfo {author} {\bibfnamefont {P.~L.}\ \bibnamefont {{Russo}}}, \bibinfo
  {author} {\bibfnamefont {A.~T.}\ \bibnamefont {{Savici}}}, \bibinfo {author}
  {\bibfnamefont {A.~A.}\ \bibnamefont {{Aczel}}}, \bibinfo {author}
  {\bibfnamefont {G.~J.}\ \bibnamefont {{MacDougall}}}, \bibinfo {author}
  {\bibfnamefont {J.~A.}\ \bibnamefont {{Rodriguez}}}, \bibinfo {author}
  {\bibfnamefont {T.~J.}\ \bibnamefont {{Williams}}}, \bibinfo {author}
  {\bibfnamefont {G.~M.}\ \bibnamefont {{Luke}}}, \bibinfo {author}
  {\bibfnamefont {C.~R.}\ \bibnamefont {{Wiebe}}}, \bibinfo {author}
  {\bibfnamefont {Y.}~\bibnamefont {{Yoshida}}}, \bibinfo {author}
  {\bibfnamefont {S.}~\bibnamefont {{Nakatsuji}}}, \bibinfo {author}
  {\bibfnamefont {Y.}~\bibnamefont {{Maeno}}}, \bibinfo {author} {\bibfnamefont
  {T.}~\bibnamefont {{Taniguchi}}}, \ and\ \bibinfo {author} {\bibfnamefont
  {Y.~J.}\ \bibnamefont {{Uemura}}},\ }\href {\doibase 10.1038/nmat3236}
  {\bibfield  {journal} {\bibinfo  {journal} {Nature Materials}\ }\textbf
  {\bibinfo {volume} {11}},\ \bibinfo {pages} {323} (\bibinfo {year}
  {2012})}\BibitemShut {NoStop}%
\bibitem [{\citenamefont {{Ortmann}}\ \emph {et~al.}(2013)\citenamefont
  {{Ortmann}}, \citenamefont {{Liu}}, \citenamefont {{Hu}}, \citenamefont
  {{Zhu}}, \citenamefont {{Peng}}, \citenamefont {{Matsuda}}, \citenamefont
  {{Ke}},\ and\ \citenamefont {{Mao}}}]{Ortmann2013sr}%
  \BibitemOpen
  \bibfield  {author} {\bibinfo {author} {\bibfnamefont {J.~E.}\ \bibnamefont
  {{Ortmann}}}, \bibinfo {author} {\bibfnamefont {J.~Y.}\ \bibnamefont
  {{Liu}}}, \bibinfo {author} {\bibfnamefont {J.}~\bibnamefont {{Hu}}},
  \bibinfo {author} {\bibfnamefont {M.}~\bibnamefont {{Zhu}}}, \bibinfo
  {author} {\bibfnamefont {J.}~\bibnamefont {{Peng}}}, \bibinfo {author}
  {\bibfnamefont {M.}~\bibnamefont {{Matsuda}}}, \bibinfo {author}
  {\bibfnamefont {X.}~\bibnamefont {{Ke}}}, \ and\ \bibinfo {author}
  {\bibfnamefont {Z.~Q.}\ \bibnamefont {{Mao}}},\ }\href {\doibase
  10.1038/srep02950} {\bibfield  {journal} {\bibinfo  {journal} {Scientific
  Reports}\ }\textbf {\bibinfo {volume} {3}},\ \bibinfo {eid} {2950} (\bibinfo
  {year} {2013}),\ 10.1038/srep02950}\BibitemShut {NoStop}%
\bibitem [{\citenamefont {Gloos}(2009)}]{Gloos2009ltp}%
  \BibitemOpen
  \bibfield  {author} {\bibinfo {author} {\bibfnamefont {K.}~\bibnamefont
  {Gloos}},\ }\href {\doibase 10.1063/1.3274810} {\bibfield  {journal}
  {\bibinfo  {journal} {Low Temperature Physics}\ }\textbf {\bibinfo {volume}
  {35}},\ \bibinfo {pages} {935} (\bibinfo {year} {2009})}\BibitemShut
  {NoStop}%
\bibitem [{\citenamefont {Gloos}\ and\ \citenamefont
  {Tuuli}(2012)}]{Gloos2012jpcs}%
  \BibitemOpen
  \bibfield  {author} {\bibinfo {author} {\bibfnamefont {K.}~\bibnamefont
  {Gloos}}\ and\ \bibinfo {author} {\bibfnamefont {E.}~\bibnamefont {Tuuli}},\
  }\href {http://stacks.iop.org/1742-6596/400/i=4/a=042011} {\bibfield
  {journal} {\bibinfo  {journal} {Journal of Physics: Conference Series}\
  }\textbf {\bibinfo {volume} {400}},\ \bibinfo {pages} {042011} (\bibinfo
  {year} {2012})}\BibitemShut {NoStop}%
\bibitem [{\citenamefont {Meekes}(1988)}]{Meekes1988prb}%
  \BibitemOpen
  \bibfield  {author} {\bibinfo {author} {\bibfnamefont {H.}~\bibnamefont
  {Meekes}},\ }\href {\doibase 10.1103/PhysRevB.38.5924} {\bibfield  {journal}
  {\bibinfo  {journal} {Phys. Rev. B}\ }\textbf {\bibinfo {volume} {38}},\
  \bibinfo {pages} {5924} (\bibinfo {year} {1988})}\BibitemShut {NoStop}%
\bibitem [{\citenamefont {Arham}\ \emph {et~al.}(2012)\citenamefont {Arham},
  \citenamefont {Hunt}, \citenamefont {Park}, \citenamefont {Gillett},
  \citenamefont {Das}, \citenamefont {Sebastian}, \citenamefont {Xu},
  \citenamefont {Wen}, \citenamefont {Lin}, \citenamefont {Li}, \citenamefont
  {Gu}, \citenamefont {Thaler}, \citenamefont {Ran}, \citenamefont {Bud'ko},
  \citenamefont {Canfield}, \citenamefont {Chung}, \citenamefont {Kanatzidis},\
  and\ \citenamefont {Greene}}]{Arham2012prb}%
  \BibitemOpen
  \bibfield  {author} {\bibinfo {author} {\bibfnamefont {H.~Z.}\ \bibnamefont
  {Arham}}, \bibinfo {author} {\bibfnamefont {C.~R.}\ \bibnamefont {Hunt}},
  \bibinfo {author} {\bibfnamefont {W.~K.}\ \bibnamefont {Park}}, \bibinfo
  {author} {\bibfnamefont {J.}~\bibnamefont {Gillett}}, \bibinfo {author}
  {\bibfnamefont {S.~D.}\ \bibnamefont {Das}}, \bibinfo {author} {\bibfnamefont
  {S.~E.}\ \bibnamefont {Sebastian}}, \bibinfo {author} {\bibfnamefont {Z.~J.}\
  \bibnamefont {Xu}}, \bibinfo {author} {\bibfnamefont {J.~S.}\ \bibnamefont
  {Wen}}, \bibinfo {author} {\bibfnamefont {Z.~W.}\ \bibnamefont {Lin}},
  \bibinfo {author} {\bibfnamefont {Q.}~\bibnamefont {Li}}, \bibinfo {author}
  {\bibfnamefont {G.}~\bibnamefont {Gu}}, \bibinfo {author} {\bibfnamefont
  {A.}~\bibnamefont {Thaler}}, \bibinfo {author} {\bibfnamefont
  {S.}~\bibnamefont {Ran}}, \bibinfo {author} {\bibfnamefont {S.~L.}\
  \bibnamefont {Bud'ko}}, \bibinfo {author} {\bibfnamefont {P.~C.}\
  \bibnamefont {Canfield}}, \bibinfo {author} {\bibfnamefont {D.~Y.}\
  \bibnamefont {Chung}}, \bibinfo {author} {\bibfnamefont {M.~G.}\ \bibnamefont
  {Kanatzidis}}, \ and\ \bibinfo {author} {\bibfnamefont {L.~H.}\ \bibnamefont
  {Greene}},\ }\href {\doibase 10.1103/PhysRevB.85.214515} {\bibfield
  {journal} {\bibinfo  {journal} {Phys. Rev. B}\ }\textbf {\bibinfo {volume}
  {85}},\ \bibinfo {pages} {214515} (\bibinfo {year} {2012})}\BibitemShut
  {NoStop}%
\bibitem [{\citenamefont {Gershenzon}\ \emph {et~al.}(1986)\citenamefont
  {Gershenzon}, \citenamefont {Gubankov},\ and\ \citenamefont
  {Fale{\v{i}}}}]{Gershenzon1986jetp}%
  \BibitemOpen
  \bibfield  {author} {\bibinfo {author} {\bibfnamefont {M.~E.}\ \bibnamefont
  {Gershenzon}}, \bibinfo {author} {\bibfnamefont {V.~N.}\ \bibnamefont
  {Gubankov}}, \ and\ \bibinfo {author} {\bibfnamefont {M.~I.}\ \bibnamefont
  {Fale{\v{i}}}},\ }\href@noop {} {\bibfield  {journal} {\bibinfo  {journal}
  {Sov. Phys. JETP}\ }\textbf {\bibinfo {volume} {63}},\ \bibinfo {pages}
  {1287} (\bibinfo {year} {1986})}\BibitemShut {NoStop}%
\bibitem [{\citenamefont {Abrikosov}(2000)}]{Abrikosov2000prb}%
  \BibitemOpen
  \bibfield  {author} {\bibinfo {author} {\bibfnamefont {A.~A.}\ \bibnamefont
  {Abrikosov}},\ }\href {\doibase 10.1103/PhysRevB.61.7770} {\bibfield
  {journal} {\bibinfo  {journal} {Phys. Rev. B}\ }\textbf {\bibinfo {volume}
  {61}},\ \bibinfo {pages} {7770} (\bibinfo {year} {2000})}\BibitemShut
  {NoStop}%
\bibitem [{\citenamefont {Mazur}\ \emph {et~al.}(2007)\citenamefont {Mazur},
  \citenamefont {Gray}, \citenamefont {Zasadzinski}, \citenamefont {Ozyuzer},
  \citenamefont {Beloborodov}, \citenamefont {Zheng},\ and\ \citenamefont
  {Mitchell}}]{Mazur2007prb}%
  \BibitemOpen
  \bibfield  {author} {\bibinfo {author} {\bibfnamefont {D.}~\bibnamefont
  {Mazur}}, \bibinfo {author} {\bibfnamefont {K.~E.}\ \bibnamefont {Gray}},
  \bibinfo {author} {\bibfnamefont {J.~F.}\ \bibnamefont {Zasadzinski}},
  \bibinfo {author} {\bibfnamefont {L.}~\bibnamefont {Ozyuzer}}, \bibinfo
  {author} {\bibfnamefont {I.~S.}\ \bibnamefont {Beloborodov}}, \bibinfo
  {author} {\bibfnamefont {H.}~\bibnamefont {Zheng}}, \ and\ \bibinfo {author}
  {\bibfnamefont {J.~F.}\ \bibnamefont {Mitchell}},\ }\href {\doibase
  10.1103/PhysRevB.76.193102} {\bibfield  {journal} {\bibinfo  {journal} {Phys.
  Rev. B}\ }\textbf {\bibinfo {volume} {76}},\ \bibinfo {pages} {193102}
  (\bibinfo {year} {2007})}\BibitemShut {NoStop}%
\bibitem [{\citenamefont {Al'tshuler}\ and\ \citenamefont
  {Aronov}(1979)}]{Altshuler1979}%
  \BibitemOpen
  \bibfield  {author} {\bibinfo {author} {\bibfnamefont {B.~L.}\ \bibnamefont
  {Al'tshuler}}\ and\ \bibinfo {author} {\bibfnamefont {A.~G.}\ \bibnamefont
  {Aronov}},\ }\href@noop {} {\bibfield  {journal} {\bibinfo  {journal} {Sov.
  Phys. JETP}\ }\textbf {\bibinfo {volume} {50}},\ \bibinfo {pages} {968}
  (\bibinfo {year} {1979})}\BibitemShut {NoStop}%
\bibitem [{\citenamefont {Altshuler}\ \emph {et~al.}(1980)\citenamefont
  {Altshuler}, \citenamefont {Aronov},\ and\ \citenamefont
  {Lee}}]{Altshuler1980}%
  \BibitemOpen
  \bibfield  {author} {\bibinfo {author} {\bibfnamefont {B.~L.}\ \bibnamefont
  {Altshuler}}, \bibinfo {author} {\bibfnamefont {A.~G.}\ \bibnamefont
  {Aronov}}, \ and\ \bibinfo {author} {\bibfnamefont {P.~A.}\ \bibnamefont
  {Lee}},\ }\href {\doibase 10.1103/PhysRevLett.44.1288} {\bibfield  {journal}
  {\bibinfo  {journal} {Phys. Rev. Lett.}\ }\textbf {\bibinfo {volume} {44}},\
  \bibinfo {pages} {1288} (\bibinfo {year} {1980})}\BibitemShut {NoStop}%
\bibitem [{\citenamefont {Altshuler}\ and\ \citenamefont
  {Aronov}(1985)}]{Altshuler1985}%
  \BibitemOpen
  \bibfield  {author} {\bibinfo {author} {\bibfnamefont {B.~L.}\ \bibnamefont
  {Altshuler}}\ and\ \bibinfo {author} {\bibfnamefont {A.~G.}\ \bibnamefont
  {Aronov}},\ }in\ \href@noop {} {\emph {\bibinfo {booktitle}
  {Electron-Electron Interactions in Disordered Systems}}},\ \bibinfo {series}
  {Modern Problems in Condensed Matter Sciences}, Vol.~\bibinfo {volume} {10},\
  \bibinfo {editor} {edited by\ \bibinfo {editor} {\bibfnamefont
  {M.}~\bibnamefont {Pollak}}\ and\ \bibinfo {editor} {\bibfnamefont {A.~L.}\
  \bibnamefont {Efros}}}\ (\bibinfo  {publisher} {North-Holland},\ \bibinfo
  {address} {Amsterdam},\ \bibinfo {year} {1985})\ Chap.~\bibinfo {chapter}
  {1}, pp.\ \bibinfo {pages} {1--154}\BibitemShut {NoStop}%
\bibitem [{\citenamefont {Liu}\ \emph {et~al.}(2014)\citenamefont {Liu},
  \citenamefont {Niu}, \citenamefont {Xiang}, \citenamefont {Wei},
  \citenamefont {Li}, \citenamefont {Feng}, \citenamefont {Han}, \citenamefont
  {Zhang},\ and\ \citenamefont {Coey}}]{Liu2014prb}%
  \BibitemOpen
  \bibfield  {author} {\bibinfo {author} {\bibfnamefont {L.}~\bibnamefont
  {Liu}}, \bibinfo {author} {\bibfnamefont {J.}~\bibnamefont {Niu}}, \bibinfo
  {author} {\bibfnamefont {L.}~\bibnamefont {Xiang}}, \bibinfo {author}
  {\bibfnamefont {J.}~\bibnamefont {Wei}}, \bibinfo {author} {\bibfnamefont
  {D.-L.}\ \bibnamefont {Li}}, \bibinfo {author} {\bibfnamefont {J.-F.}\
  \bibnamefont {Feng}}, \bibinfo {author} {\bibfnamefont {X.-F.}\ \bibnamefont
  {Han}}, \bibinfo {author} {\bibfnamefont {X.-G.}\ \bibnamefont {Zhang}}, \
  and\ \bibinfo {author} {\bibfnamefont {J.~M.~D.}\ \bibnamefont {Coey}},\
  }\href {\doibase 10.1103/PhysRevB.90.195132} {\bibfield  {journal} {\bibinfo
  {journal} {Phys. Rev. B}\ }\textbf {\bibinfo {volume} {90}},\ \bibinfo
  {pages} {195132} (\bibinfo {year} {2014})}\BibitemShut {NoStop}%
\bibitem [{\citenamefont {Sheet}\ \emph {et~al.}(2004)\citenamefont {Sheet},
  \citenamefont {Mukhopadhyay},\ and\ \citenamefont
  {Raychaudhuri}}]{Sheet2004prb}%
  \BibitemOpen
  \bibfield  {author} {\bibinfo {author} {\bibfnamefont {G.}~\bibnamefont
  {Sheet}}, \bibinfo {author} {\bibfnamefont {S.}~\bibnamefont {Mukhopadhyay}},
  \ and\ \bibinfo {author} {\bibfnamefont {P.}~\bibnamefont {Raychaudhuri}},\
  }\href {\doibase 10.1103/PhysRevB.69.134507} {\bibfield  {journal} {\bibinfo
  {journal} {Phys. Rev. B}\ }\textbf {\bibinfo {volume} {69}},\ \bibinfo
  {pages} {134507} (\bibinfo {year} {2004})}\BibitemShut {NoStop}%
\bibitem [{\citenamefont {Duif}\ \emph {et~al.}(1989)\citenamefont {Duif},
  \citenamefont {Jansen},\ and\ \citenamefont {Wyder}}]{Duif1989jpcm}%
  \BibitemOpen
  \bibfield  {author} {\bibinfo {author} {\bibfnamefont {A.~M.}\ \bibnamefont
  {Duif}}, \bibinfo {author} {\bibfnamefont {A.~G.~M.}\ \bibnamefont {Jansen}},
  \ and\ \bibinfo {author} {\bibfnamefont {P.}~\bibnamefont {Wyder}},\ }\href
  {http://stacks.iop.org/0953-8984/1/i=20/a=001} {\bibfield  {journal}
  {\bibinfo  {journal} {Journal of Physics: Condensed Matter}\ }\textbf
  {\bibinfo {volume} {1}},\ \bibinfo {pages} {3157} (\bibinfo {year}
  {1989})}\BibitemShut {NoStop}%
\bibitem [{\citenamefont {Naidyuk}\ and\ \citenamefont
  {Yanson}(1998)}]{Naidyuk1998jpcm}%
  \BibitemOpen
  \bibfield  {author} {\bibinfo {author} {\bibfnamefont {Y.~G.}\ \bibnamefont
  {Naidyuk}}\ and\ \bibinfo {author} {\bibfnamefont {I.~K.}\ \bibnamefont
  {Yanson}},\ }\href {http://stacks.iop.org/0953-8984/10/i=40/a=001} {\bibfield
   {journal} {\bibinfo  {journal} {Journal of Physics: Condensed Matter}\
  }\textbf {\bibinfo {volume} {10}},\ \bibinfo {pages} {8905} (\bibinfo {year}
  {1998})}\BibitemShut {NoStop}%
\bibitem [{\citenamefont {Park}\ and\ \citenamefont
  {Greene}(2009)}]{Park2009jpcm}%
  \BibitemOpen
  \bibfield  {author} {\bibinfo {author} {\bibfnamefont {W.~K.}\ \bibnamefont
  {Park}}\ and\ \bibinfo {author} {\bibfnamefont {L.~H.}\ \bibnamefont
  {Greene}},\ }\href {http://stacks.iop.org/0953-8984/21/i=10/a=103203}
  {\bibfield  {journal} {\bibinfo  {journal} {Journal of Physics: Condensed
  Matter}\ }\textbf {\bibinfo {volume} {21}},\ \bibinfo {pages} {103203}
  (\bibinfo {year} {2009})}\BibitemShut {NoStop}%
\bibitem [{\citenamefont {Pothier}\ \emph {et~al.}(1997)\citenamefont
  {Pothier}, \citenamefont {Gueron}, \citenamefont {Birge}, \citenamefont
  {Esteve},\ and\ \citenamefont {Devoret}}]{Pothier1997}%
  \BibitemOpen
  \bibfield  {author} {\bibinfo {author} {\bibfnamefont {H.}~\bibnamefont
  {Pothier}}, \bibinfo {author} {\bibfnamefont {S.}~\bibnamefont {Gueron}},
  \bibinfo {author} {\bibfnamefont {N.~O.}\ \bibnamefont {Birge}}, \bibinfo
  {author} {\bibfnamefont {D.}~\bibnamefont {Esteve}}, \ and\ \bibinfo {author}
  {\bibfnamefont {M.~H.}\ \bibnamefont {Devoret}},\ }\href@noop {} {\bibfield
  {journal} {\bibinfo  {journal} {Phys. Rev. Lett.}\ }\textbf {\bibinfo
  {volume} {79}},\ \bibinfo {pages} {3490} (\bibinfo {year}
  {1997})}\BibitemShut {NoStop}%
\bibitem [{\citenamefont {Blonder}\ \emph {et~al.}(1982)\citenamefont
  {Blonder}, \citenamefont {Tinkham},\ and\ \citenamefont
  {Klapwijk}}]{Blonder1982prb}%
  \BibitemOpen
  \bibfield  {author} {\bibinfo {author} {\bibfnamefont {G.~E.}\ \bibnamefont
  {Blonder}}, \bibinfo {author} {\bibfnamefont {M.}~\bibnamefont {Tinkham}}, \
  and\ \bibinfo {author} {\bibfnamefont {T.~M.}\ \bibnamefont {Klapwijk}},\
  }\href {\doibase 10.1103/PhysRevB.25.4515} {\bibfield  {journal} {\bibinfo
  {journal} {Phys. Rev. B}\ }\textbf {\bibinfo {volume} {25}},\ \bibinfo
  {pages} {4515} (\bibinfo {year} {1982})}\BibitemShut {NoStop}%
\bibitem [{\citenamefont {Peng}\ \emph {et~al.}(2013)\citenamefont {Peng},
  \citenamefont {De}, \citenamefont {Lv}, \citenamefont {Wei},\ and\
  \citenamefont {Chu}}]{Peng2013prb}%
  \BibitemOpen
  \bibfield  {author} {\bibinfo {author} {\bibfnamefont {H.}~\bibnamefont
  {Peng}}, \bibinfo {author} {\bibfnamefont {D.}~\bibnamefont {De}}, \bibinfo
  {author} {\bibfnamefont {B.}~\bibnamefont {Lv}}, \bibinfo {author}
  {\bibfnamefont {F.}~\bibnamefont {Wei}}, \ and\ \bibinfo {author}
  {\bibfnamefont {C.-W.}\ \bibnamefont {Chu}},\ }\href {\doibase
  10.1103/PhysRevB.88.024515} {\bibfield  {journal} {\bibinfo  {journal} {Phys.
  Rev. B}\ }\textbf {\bibinfo {volume} {88}},\ \bibinfo {pages} {024515}
  (\bibinfo {year} {2013})}\BibitemShut {NoStop}%
\bibitem [{\citenamefont {Bugoslavsky}\ \emph {et~al.}(2005)\citenamefont
  {Bugoslavsky}, \citenamefont {Miyoshi}, \citenamefont {Perkins},
  \citenamefont {Caplin}, \citenamefont {Cohen}, \citenamefont {Pogrebnyakov},\
  and\ \citenamefont {Xi}}]{Bugoslavsky2005prb}%
  \BibitemOpen
  \bibfield  {author} {\bibinfo {author} {\bibfnamefont {Y.}~\bibnamefont
  {Bugoslavsky}}, \bibinfo {author} {\bibfnamefont {Y.}~\bibnamefont
  {Miyoshi}}, \bibinfo {author} {\bibfnamefont {G.~K.}\ \bibnamefont
  {Perkins}}, \bibinfo {author} {\bibfnamefont {A.~D.}\ \bibnamefont {Caplin}},
  \bibinfo {author} {\bibfnamefont {L.~F.}\ \bibnamefont {Cohen}}, \bibinfo
  {author} {\bibfnamefont {A.~V.}\ \bibnamefont {Pogrebnyakov}}, \ and\
  \bibinfo {author} {\bibfnamefont {X.~X.}\ \bibnamefont {Xi}},\ }\href
  {\doibase 10.1103/PhysRevB.72.224506} {\bibfield  {journal} {\bibinfo
  {journal} {Phys. Rev. B}\ }\textbf {\bibinfo {volume} {72}},\ \bibinfo
  {pages} {224506} (\bibinfo {year} {2005})}\BibitemShut {NoStop}%
\bibitem [{\citenamefont {Daghero}\ \emph {et~al.}(2013)\citenamefont
  {Daghero}, \citenamefont {Tortello}, \citenamefont {Pecchio}, \citenamefont
  {Stepanov},\ and\ \citenamefont {Gonnelli}}]{Daghero2013ltp}%
  \BibitemOpen
  \bibfield  {author} {\bibinfo {author} {\bibfnamefont {D.}~\bibnamefont
  {Daghero}}, \bibinfo {author} {\bibfnamefont {M.}~\bibnamefont {Tortello}},
  \bibinfo {author} {\bibfnamefont {P.}~\bibnamefont {Pecchio}}, \bibinfo
  {author} {\bibfnamefont {V.~A.}\ \bibnamefont {Stepanov}}, \ and\ \bibinfo
  {author} {\bibfnamefont {R.~S.}\ \bibnamefont {Gonnelli}},\ }\href@noop {}
  {\bibfield  {journal} {\bibinfo  {journal} {Low Temperature Physics}\
  }\textbf {\bibinfo {volume} {39}} (\bibinfo {year} {2013})}\BibitemShut
  {NoStop}%
\bibitem [{\citenamefont {Lichtenberg}(2002)}]{Lichtenberg2002pssc}%
  \BibitemOpen
  \bibfield  {author} {\bibinfo {author} {\bibfnamefont {F.}~\bibnamefont
  {Lichtenberg}},\ }\href {\doibase 10.1016/j.progsolidstchem.2003.07.001}
  {\bibfield  {journal} {\bibinfo  {journal} {Progress in Solid State
  Chemistry}\ }\textbf {\bibinfo {volume} {30}},\ \bibinfo {pages} {103 }
  (\bibinfo {year} {2002})}\BibitemShut {NoStop}%
\bibitem [{\citenamefont {Sasaki}\ \emph {et~al.}(2011)\citenamefont {Sasaki},
  \citenamefont {Kriener}, \citenamefont {Segawa}, \citenamefont {Yada},
  \citenamefont {Tanaka}, \citenamefont {Sato},\ and\ \citenamefont
  {Ando}}]{Sasaki2011prl}%
  \BibitemOpen
  \bibfield  {author} {\bibinfo {author} {\bibfnamefont {S.}~\bibnamefont
  {Sasaki}}, \bibinfo {author} {\bibfnamefont {M.}~\bibnamefont {Kriener}},
  \bibinfo {author} {\bibfnamefont {K.}~\bibnamefont {Segawa}}, \bibinfo
  {author} {\bibfnamefont {K.}~\bibnamefont {Yada}}, \bibinfo {author}
  {\bibfnamefont {Y.}~\bibnamefont {Tanaka}}, \bibinfo {author} {\bibfnamefont
  {M.}~\bibnamefont {Sato}}, \ and\ \bibinfo {author} {\bibfnamefont
  {Y.}~\bibnamefont {Ando}},\ }\href {\doibase 10.1103/PhysRevLett.107.217001}
  {\bibfield  {journal} {\bibinfo  {journal} {Phys. Rev. Lett.}\ }\textbf
  {\bibinfo {volume} {107}},\ \bibinfo {pages} {217001} (\bibinfo {year}
  {2011})}\BibitemShut {NoStop}%
\end{thebibliography}

%

\end{document}